\documentclass[acmsmall, screen]{acmart}
\AtBeginDocument{%
  }

\newcommand{\eg}{\textit{e.g., }}
\newcommand{\ie}{\textit{i.e., }}
\newcommand{\etal}{et~al. }

\newcommand{\myverb}[1]{%
    \unskip%
    \mbox{%
        \fontsize{10}{12}
        \usefont{OT1}{lmtt}{b}{n}%
        \spaceskip=0.25em\relax 
        #1%
    }%
}

\graphicspath{Figs/}
\DeclareGraphicsExtensions{.pdf,.jpeg,.png}

\usepackage{textcomp}

\begin{document}

\title[Mapping Digital Systems with the Bill of Materials]{A Survey on Mapping Digital Systems with Bill of Materials: Development, Practices, and Challenges}

\author{Shuai~Zhang}
\orcid{0009-0009-1288-5460}
\affiliation{%
  \institution{University of New South Wales}
  \city{Sydney}
  \state{NSW}
  \country{Australia}
}
\email{shuai.zhang4@student.unsw.edu.au}

\author{Minzhao~Lyu}
\orcid{0000-0001-8677-248X}
\affiliation{%
  \institution{University of New South Wales}
  \city{Sydney}
  \state{NSW}
  \country{Australia}
}
\email{minzhao.lyu@unsw.edu.au}

\author{Hassan Habibi~Gharakheili}
\orcid{0000-0002-9333-7635}
\affiliation{%
  \institution{University of New South Wales}
  \city{Sydney}
  \state{NSW}
  \country{Australia}
}
\email{h.habibi@unsw.edu.au}

\begin{abstract}

Modern digital ecosystems, spanning software, hardware, learning models, datasets, and cryptographic products, continue to grow in complexity, making it difficult for organizations to understand and manage component dependencies. Bills of Materials (BOMs) have emerged as a structured way to document product components, their interrelationships, and key metadata, improving visibility and security across digital supply chains.
This survey provides the first comprehensive cross-domain review of BOM developments and practices.  
We start by examining the evolution of BOM frameworks in three stages (\ie pre-development, initial, and accelerated) and summarizing their core principles, key stakeholders, and standardization efforts for hardware, software, artificial intelligence (AI) models, datasets, and cryptographic assets.
We then review industry practices for generating BOM data, evaluating its quality, and securely sharing it. 
Next, we review practical downstream uses of BOM data, including dependency modeling, compliance verification, operational risk assessment, and vulnerability tracking.
We also discuss academic efforts to address limitations in current BOM frameworks through refinements, extensions, or new models tailored to emerging domains such as data ecosystems and AI supply chains.
Finally, we identify four key gaps that limit the usability and reliability of today's BOM frameworks, motivating future research directions.

\end{abstract}




\maketitle

\section{Introduction}
Modern digital ecosystems are becoming increasingly interconnected, driven in part by Industry 4.0 and the broader digitization of organizational workflows. Today’s systems integrate software, firmware, cloud services, Internet of Things (IoT) devices, artificial intelligence (AI) models, and large-scale data pipelines, creating complex and deeply dependent structures. Because these digital components directly influence system behavior and data flows, they have become critical to ensuring reliability, maintaining operational continuity, and protecting against cyber threats in the supply chain \cite{innovapptive:industryFour2025}.

Despite these advances, organizations often lack visibility into the dependencies of the components that make up their digital supply chains. This limited transparency creates hidden cyber risks with potentially significant economic and safety impacts. 
The World Economic Forum's recent global cybersecurity insight report \cite{weforum:cybersecurityOutlook2025} found that 72\% of organizations experienced major cyber incidents in the past year, and 54\% of large organizations identified poor visibility into component dependencies as a key cause, particularly for cloud infrastructures, cyber-physical systems, and AI supply chains. 
A widely-cited example is the Log4Shell vulnerability (\ie CVE-2021-44228), disclosed in December 2021, which was embedded as a hidden dependency in more than 10 million software projects, propagating rapidly through cloud services, web servers, enterprise software, and mobile applications in sectors like retail, technology, finance, manufacturing, and healthcare \cite{upguard:log4jVulnerability2025}. 

To address these challenges, the software community introduced the Software Bill of Materials (SBOM), a structured inventory that lists the components, versions, licenses, and dependency relationships of a software product. SBOMs have been advanced through industry initiatives such as OWASP and the Linux Foundation, developed through standards like CycloneDX and SPDX, and formally defined by the U.S. National Telecommunications and Information Administration (NTIA) \cite{ntia:sbomMinimumElements2021}. 
Governments around the world have begun to require SBOMs to improve supply chain security and transparency. In the United States, Executive Order 14028 \cite{federalregister:executiveOrder2021} mandates SBOMs for software sold to federal agencies, accelerating industry adoption. In Europe, the Cyber Resilience Act (CRA) \cite{europa:cyberResilienceAct2024} requires ongoing SBOM updates throughout a software product's lifecycle.
Beyond software, BOM frameworks now extend to hardware (HBOM), cloud service (SaaSBOM), AI models and datasets (AIBOM), and cryptographic systems (CBOM), reflecting broader industry and regulatory adoption \cite{cisa:hbomFramework2023, cisa:saasTransparency2024, KBennetLinuxFoundation2024, cyclonedx:cbomAuthoritativeGuide2024}.

\textbf{\textit{Survey Scope and Methodology:}} 
Despite the growing use of BOM across software, hardware, AI, data, cloud services, and cryptographic systems, current efforts remain fragmented, leaving practitioners without a unified understanding of how these different BOM frameworks relate or complement one another.  
This survey addresses this gap by providing the first cross-domain review of BOM developments, practices, and research published since 2020, just one year before BOM standards were formalized. We cover the evolution of BOM standards and industry adoption, scientific advances in BOM generation and management, practical applications of BOM data for security and transparency, and recent extensions designed to address current limitations.
To ensure broad coverage, we conducted a systematic search on Google Scholar, ACM Digital Library, IEEE Xplore, and arXiv. Our searches used keyword combinations targeting five main BOM categories: SBOM (\eg keywords including ``sbom'' OR ``software bill of material'' OR ``cyclonedx'' OR ``software package data exchange'' OR ``spdx'' are used to search for SBOM-related papers), CBOM, AIBOM, HBOM, and SaaSBOM. 
The surveyed works span the leading venues in software engineering (\eg FSE, ICSE, ISSTA), cybersecurity (\eg IEEE S\&P, NDSS), and supply chain security (\eg ACM SCORED). We also incorporate relevant Internet standards, industry reports, and research addressing broader challenges in digital supply chain transparency.

\begin{figure*}[t]
    \centering
    \captionsetup{skip=2mm}
    \includegraphics[width=0.995\linewidth]{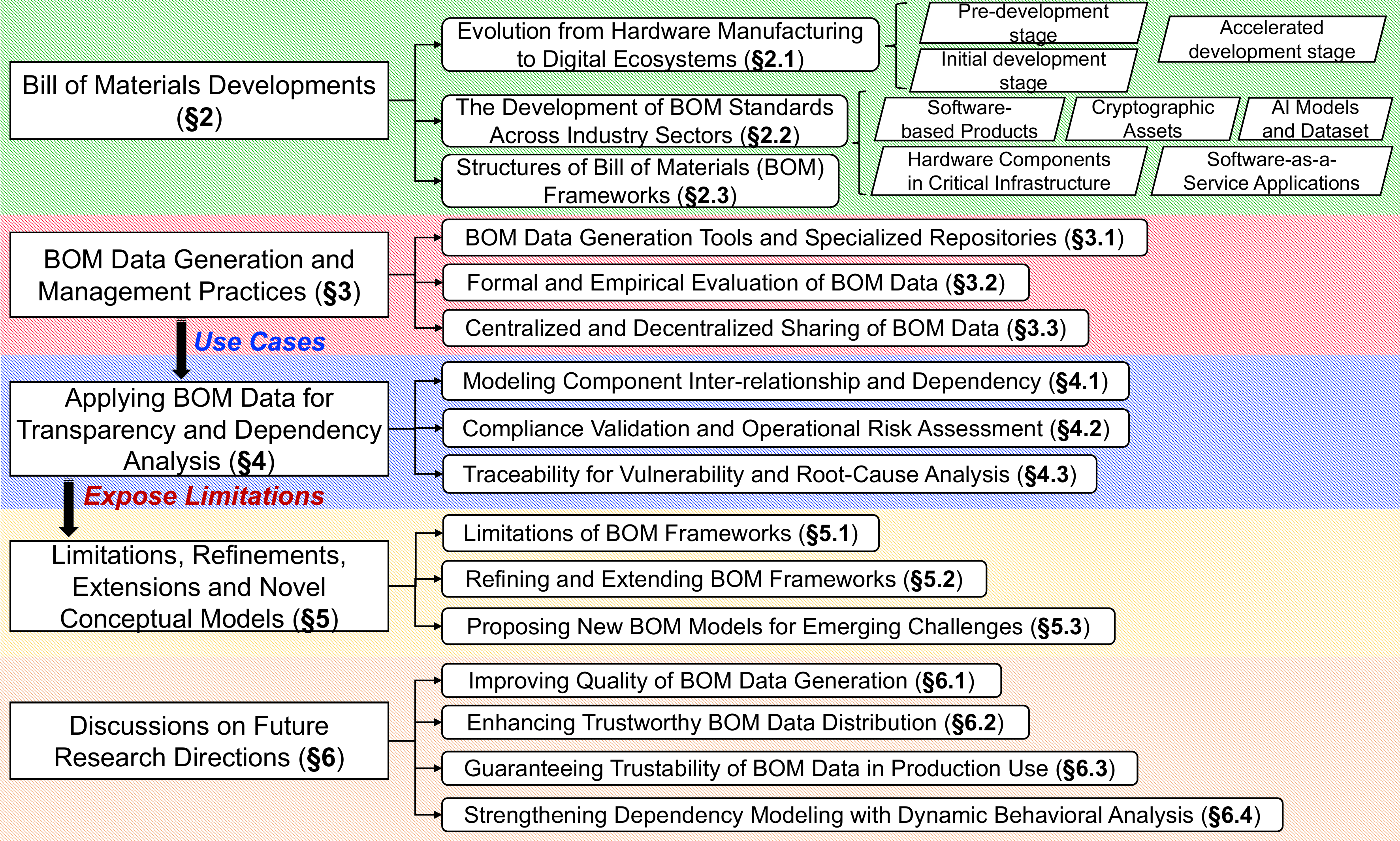}
    \caption{Key topics covered in this survey.}
    \label{fig:roadmap}
    \vspace{-5mm}
\end{figure*}

\textbf{\textit{Roadmap:}} 
The roadmap of this article is illustrated in Fig. \ref{fig:roadmap}. 
We start by reviewing the three-stage development of the BOM framework in \S\ref{sec:bom_developments}, highlighting its evolution from traditional manufacturing blueprints to comprehensive digital component inventories.
Next, in \S\ref{sec:bom_generation_management}, we examine industry and academic practices throughout the lifecycle of the BOM, including efforts to improve the quality of BOM data, assess usability, and implement reliable mechanisms for sharing BOM data while maintaining integrity. 
In \S\ref{sec:bom_application}, we explore practical use cases of BOM data, such as constructing inventories of product components and dependencies, verifying compliance, evaluating operational risks, identifying vulnerabilities, and tracing root causes.
After that, \S\ref{sec:bom_extension} surveys the research addressing the limitations of current BOM frameworks (\eg insufficient coverage for emerging ecosystems like AI), and presents more refined, extended, or new BOM models for domains, such as AI systems, datasets, dynamic deployments, and cyber-physical/IoT systems. 
In \S\ref{sec:bom_future_direction}, we outline four future research directions to improve the usability and reliability of BOM frameworks. 
Related surveys on improving transparency in digital supply chains are discussed in \S\ref{sec:bom_related_surveys}, and the article concludes in \S\ref{sec:bom_survey_conclusion}.

\section{Bill of Materials Developments}
\label{sec:bom_developments}
A Bill of Materials (BOM) is a structured representation of the components that make up a product, along with the relationships between those components. Traditionally used in manufacturing to track physical parts, the BOM concept has evolved to encompass digital products such as software, hardware, machine learning (ML) or artificial intelligence (AI) models, datasets, and cloud services. This evolution indicates the growing need for transparency, traceability, and accountability in increasingly complex supply chains.  
In this section, we first describe the historical development of BOMs in different domains and time periods (\S\ref{subsec:bom_evolution}). We then provide domain-specific discussions of BOM practices (\S\ref{subsec:bom_dev_domains}), followed by a comparative analysis of their structural characteristics (\S\ref{subsec:bom_structure_analysis}).

\subsection{Evolution from Hardware Manufacturing to Digital Ecosystems}
\label{subsec:bom_evolution}
\begin{figure*}[t]
    \centering
    \captionsetup{skip=2mm}
    \includegraphics[width=0.995\linewidth]{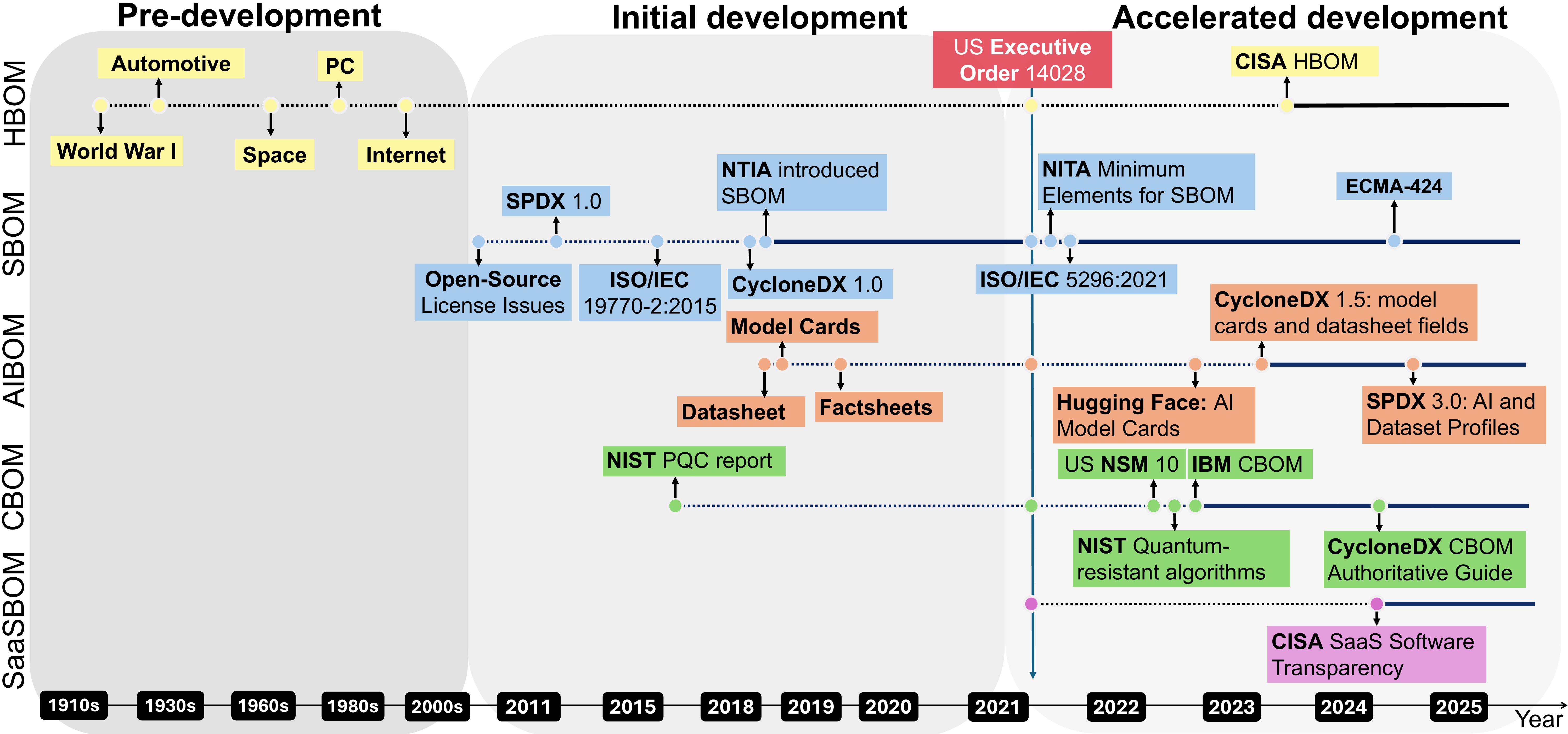}
    \caption{Development of BOM concepts across three stages (1910s---2025), moving from hardware-focused use-cases to software, AI/data, cryptographic, and SaaS-specific implementations. Dashed lines indicate pre-standardization developments, and solid lines represent periods of formalized BOM standards.}
    \label{fig:bom_development}
    \vspace{-5mm}
\end{figure*}

To understand the development of BOM, we trace its roots from traditional hardware manufacturing to its current role in digital ecosystems. Over time, BOM has expanded from documenting physical components in industrial production to capturing the complex interdependencies of software, datasets, cloud services, and other digital assets. 
To better contextualize this evolution, we categorize its historical trajectory into three stages: pre-development, initial development, and accelerated development, as shown in Fig.~\ref{fig:bom_development}. Solid lines represent periods of formalized BOM standards, while dashed lines indicate pre-standardization developments.

\subsubsection{Pre-development stage}
The earliest form of BOM emerged during World War I (1914-1918) in the form of engineering drawings that itemized components and their quantities for weapon manufacturing \cite{netsuite:BoM2025, freepage:wwiWeapons2003}. At the same time, the U.S. War Industries Board (WIB) pursued standardization efforts, such as the .30-06 Springfield ammo that could interchange across multiple rifle and machine-gun platforms \cite{wikipedia:springField2025}), which established the technical prerequisites for mass production. These practices laid the groundwork for systematic material documentation and the foundations for modern BOM development \cite{GAsnerSSRN1993}.

In the 1920s-1930s, the automotive industry advanced these principles. Although formal BOM documentation was not standardized, manufacturers such as Ford and GM relied on structured parts lists to manage components and assembly tasks \cite{billofrightsinstiture:fordIndustrialization}. This approach enabled economies of scale, allowing the major automakers (\ie Ford, GM, and Chrysler) to dominate nearly 80\% of U.S. industry output by the late 1930s \cite{etlsolutions:automotive2020}.

A major leap occurred in the 1960s with the advent of computer technology. IBM's Information Management Systems (IMS) \cite{wikipedia:engineeringBoM2025}, originally designed to track millions of parts for NASA's Apollo spacecraft program \cite{ibm:IMS2025, wikipedia:ibmIMS2025}, transformed manual lists into hierarchical data structures capable of handling versioning and engineering changes. This marked the beginning of digital BOM management.

The 1980s personal computer revolution highlighted divergent BOM philosophies. The IBM PC (1981), built with an open and modular design, allowed third parties to create compatible hardware and software, helping global supply chains grow \cite{wikipedia:ibmPC2025, carleton:ibmVintagePC}.
In contrast, Apple's Macintosh (1984) used a closed design that limited hardware changes to protect the user experience \cite{appleinsider:Macintosh2024}.
This difference between open and closed systems showed how design choices could influence a company's market position and the growth of its ecosystem.
In the 1990s, BOM systems evolved into networked digital infrastructures.
Dell's Build-to-Order (BTO) model allowed customers to build PCs online, transforming the BOM into a dynamic, customer-driven manufacturing tool instead of a static inventory list \cite{brie:dellPC2003}. At the same time, the wide adoption of Enterprise Resource Planning (ERP) systems integrated BOM management into broader enterprise workflows \cite{wikipedia:erp2025}, enabling mass customization and global supply chain coordination.
As depicted in Fig.~\ref{fig:bom_development} (top left), the pre-development stage is dominated by Hardware BOMs (HBOM), laying the foundation for the subsequent evolution toward other BOM systems.

\subsubsection{Initial development stage}
From the 1990s to the 2010s, advances in PC hardware significantly accelerated the growth of software ecosystems. Operating systems such as Windows and Linux evolved rapidly, and competition between web browsers increased, indicating the increasing complexity of software applications \cite{stacklegend:itIndustryStories2025}. 
At the same time, open-source libraries and projects grew exponentially, driven by increased computing power (following Moore's Law) and wider access for developers \cite{dev:hardwareEvolution2024, linkedin:softwareEvolution2024}. As software tools and applications increasingly reused these open-source components, challenges around licensing compliance \cite{industryweek:openSourceLicenseCompliance2010} and security \cite{contentstack:openSourceRisk2017} emerged. These pressures drove the early adoption of BOM concepts in the software industry, which extend beyond traditional hardware to include software packages, AI models, datasets, and cryptographic assets, as shown in the middle of Fig.~\ref{fig:bom_development}.

\subsubsection{Accelerated development stage}
In May 2021, the Biden administration's Executive Order 14028, Improving the Nation's Cybersecurity \cite{federalregister:executiveOrder2021}, marked a turning point in BOM development by mandating stronger security for software supply chains. Since then, BOM concepts have expanded rapidly across sectors, driven by the need for transparency, accountability, and resilience. This accelerated phase (shown on the right side of Fig.~\ref{fig:bom_development}) set the stage for the emergence of standardized BOMs, which we discuss in the next section.

In parallel, regulators and industry leaders also advanced the transparency of the digital supply chain through efforts not directly tied to BOM frameworks. For example, Google’s SLSA framework in 2021 \cite{googleblog:slsaFramework2021}, now under OpenSSF, defines graded assurance levels for the integrity of software artifacts. Its core idea is build provenance \cite{slsa:slsaFramework2025}: verifiable metadata showing how an artifact was built, by whom, and with what inputs. The four levels of SLSA (L0–L3) range from non-provenance to hardened, tamper-resistant build systems.
NIST followed in 2022 with the Secure Software Development Framework (SSDF) \cite{nist:ssdfFramework2022}, a high-level set of practices (\eg organizational training, validating third-party components, and conducting code reviews), intended to reduce vulnerabilities and support more consistent communication between software producers and buyers.

As AI systems became widely deployed, regulators extended these efforts. 
In July 2024, NIST released SP 800-218A as a Community Profile \cite{nist:ssdfAIFramework2024} to adapt SSDF for AI, adding guidance on dataset integrity, adversarial testing, provenance checks, and mitigation of model-specific threats like theft or prompt injection.
The EU's Artificial Intelligence Act, the first comprehensive AI regulatory framework effective from August 2024 \cite{nist:ssdfAIFramework2024}, applies a risk-based model with four categories from unacceptable in social scoring practices to lower-risk systems like chatbots and sets corresponding obligations. It also regulates general-purpose AI (GPAI) models, which require transparency, documentation, testing, and systemic-risk controls for high-impact models.
The EU's Digital Product Passport (DPP), under the Ecodesign for Sustainable Products Regulation (ESPR) in 2024, similarly advances supply chain traceability \cite{wise:euDigitalProductPassport2025} by requiring manufacturers to provide machine-readable lifecycle data covering materials, environmental impact, repair instructions, and recycling pathways, accessible for 5-10 years after a product leaves the market.

\subsection{The Development of BOM Standards Across Industry Sectors}
\label{subsec:bom_dev_domains}

As shown in Fig.~\ref{fig:bom_development}, since 2011, standard bodies and industry stakeholders have worked to define BOM specifications from the early stages of development. The idea started in the software industry (\S\ref{subsubsec:software_products}), and later expanded to cover cryptographic assets (\S\ref{subsubsec:cryptographic_assets}), AI products (\S\ref{subsubsec:ai_models_data}), hardware components (\S\ref{subsubsec:hardware_compositions}), and software-as-a-service ecosystems (\S\ref{subsubsec:saas_software}).

\subsubsection{Software-based Products}
\label{subsubsec:software_products}
The software industry experienced significant growth around 2011, marked by a shift from standard self-contained software product to custom-developed applications \cite{forrester:softwareMarketTransformation2011}. This transformation encouraged more flexible and agile software development practices by leveraging existing licensed modules (each providing specific capabilities), allowing software products and systems to achieve greater functional complexity while remaining cost-effective and commercially viable. 

To effectively map the dependencies of software products that incorporate numerous licensed modules as building blocks, the first version of the Software Package Data Exchange (SPDX) standard was released in August 2011 as a community-driven effort. SPDX provides a structured framework for describing detailed information about software packages and modules, such as version numbers, names, and brief descriptions, along with their licenses, in a machine-readable format (\eg XML or Tag:Value) \cite{linuxfound:spdxRelease2010}. 

As shown in the third blue box in the SBOM thread in Fig.~\ref{fig:bom_development}, the need for the structured identification of licensed software components has grown over the years. This led to the standardization of Software Identification Tags (SWID) by the International Standard Organization (ISO/IEC) in October 2015, which record key identifying information (\eg product name, version, unique ID, creator name, licenses) for software modules using the standard XML data format \cite{nist:swidTagGuidelines2016}.
Driven by software security needs, the OWASP Foundation released the first version of CycloneDX in March 2018 (\ie the fourth blue box of the SBOM row in Fig.~\ref{fig:bom_development}), which describes software components in structured format (\eg XML) with the objective of assessing software vulnerabilities propagated through the dependent components identified by CPE (Common Platform Enumeration) URIs \cite{github:cycloneDXSpecification2024, cyclonedx:xmlReference2018}.
As shown in the fifth blue box of the SBOM thread in Fig.~\ref{fig:bom_development}, in July 2018, the National Telecommunications and Information Administration (NITA) formalized the concept of Software Bill of Materials (SBOM) to support systematic dependency modeling in the software industry \cite{wileyconnect:ntiaSBOMInitiate2018}.
NTIA later defined the minimum required SBOM elements \cite{ntia:sbomMinimumElements2021}, after the U.S. federal mandate to strengthen the security of the software supply chain \cite{federalregister:executiveOrder2021} in 2021 (the sixth and seventh blue boxes of the SBOM thread in Fig.~\ref{fig:bom_development}). In September of the same year, SPDX became the first international standard of SBOM (ISO/IEC 5296:2021 \cite{spdx:spdxISOStandard2021}), as shown in the blue box on the second right of the SBOM thread in Fig.~\ref{fig:bom_development}. In June 2024, CycloneDX 1.6 was also approved as an international SBOM standard under ECMA-424 \cite{ecma:cyclonedxECMAStandard2024} (\ie the rightmost blue box of the SBOM thread in Fig.~\ref{fig:bom_development}).

\subsubsection{Cryptographic Assets}
\label{subsubsec:cryptographic_assets}

With the proof-of-concept of quantum computing gradually validated by industry and academia, such as IBM's public 5-qubit quantum processor on the IBM Quantum Experience platform \cite{ibm:cloudQuantumComputing2016}, and the silicon-based quantum chip demonstrated by the University of New South Wales \cite{unsw:quantumDesign2016}, as shown in the first two green boxes of the CBOM thread in Fig.~\ref{fig:bom_development}, respectively, government agencies started assessing its impact on existing cryptographic systems. In 2016, NIST warned that widely used encryption algorithms (\eg RSA and ECDH) are vulnerable to quantum attacks in its post-quantum cryptography report \cite{LChenNIST2016} (\ie the third green box of the CBOM thread in Fig.~\ref{fig:bom_development}). As shown in the fourth and fifth green boxes of the CBOM thread in Fig.~\ref{fig:bom_development}, these vulnerabilities were later recognized as national security risks in the U.S. through Executive Order 14028 \cite{federalregister:executiveOrder2021} and National Security Memorandum 10 (NSM-10) \cite{fas:nationalSecurityMemo2022} issued in 2021 and 2022, respectively. In response, NIST recommended four quantum-resistant algorithms (including CRYSTALS-KYBER, CRYSTALS-Dilithium, FALCON, and SPHINCS+) for future standardization and policy adoption \cite{nist:pqgStandardized2022}, as shown in the sixth green box of the CBOM thread in Fig.~\ref{fig:bom_development}.

To support vulnerability analysis of cryptographic dependencies in digital systems, IBM proposed the first Cryptographic Bill of Materials (CBOM) concept in December 2022, extending the existing BOM originally designed for software projects \cite{github:ibmCBOM2022} (\ie the second right green box of the CBOM thread in Fig.~\ref{fig:bom_development}). 
The proposal introduced new data fields to describe cryptographic elements, such as algorithms, key management systems, and security protocols. CBOM gained broader acceptance in April 2024, when the OWASP Foundation released the Authoritative Guide to CBOM \cite{cyclonedx:cbomAuthoritativeGuide2024}, which formally documented its structure, required fields, and usage guidelines, as shown in the rightmost green box of the CBOM thread in Fig.~\ref{fig:bom_development}.

\subsubsection{AI Models and Dataset}
\label{subsubsec:ai_models_data}
For the emerging AI domain, modeling the dependencies of AI systems, such as underlying model architectures, hyperparameters, training processes, and datasets, has become increasingly important to provide stakeholders with transparency into model functionality and limitations. 
Early community efforts introduced the concept of ``Datasheet for Datasets'' \cite{TGebruarXiv2018} and ``Model Cards for Model Reporting'' \cite{MMitchellarXiv2018} in 2018 (\ie the first and second orange boxes of the AIBOX thread in Fig.~\ref{fig:bom_development}, respectively) to document key dataset and model attributes. In 2019, Arnoldar \etal \cite{MArnoldarXiv2019} further proposed a structured documentation format for AI services that includes service functionalities, model details, dataset information, and security checks, as shown in the third orange box of the AIBOM thread in Fig.~\ref{fig:bom_development}.

As shown in the fourth orange box of the AIBOM thread in Fig.~\ref{fig:bom_development}, these community initiatives were later acknowledged in the Executive Order 14028 \cite{federalregister:executiveOrder2021}. Hugging Face, a major open-source AI platform, adopted model cards \cite{huggingface:modelCards2022} to document the model specifications and intended uses (\ie the fifth orange box of the AIBOM thread in Fig.~\ref{fig:bom_development}).
Building on this momentum, BOM standards started extending support to AI components. 
In June 2023, CycloneDX added dedicated ``modelCard'' and ``data'' fields in version 1.5
\cite{cyclonedx:jsonReference2023}, forming its AIBOM specification, and later in October 2024, SPDX, introduced similar AI-focused capabilities in version 3.0 through specialized profiles for AI models and datasets \cite{KBennetLinuxFoundation2024, spdx:spdxSpecification2024}, as shown in the second right and rightmost orange boxes of the AIBOM thread in Fig.~\ref{fig:bom_development}, respectively.

\subsubsection{Hardware Components in Critical Infrastructure}
\label{subsubsec:hardware_compositions}

Although hardware supply chains have existed for more than a century, they have become increasingly complex and susceptible to cascading risks from compromised components, especially within critical infrastructures and advanced manufacturing sectors \cite{wtwco:supplyChainReport2023}. The U.S. government has highlighted these vulnerabilities in the Executive Order 14028 \cite{federalregister:executiveOrder2021}, as shown in the second right yellow box of the HBOM thread in Fig.~\ref{fig:bom_development}, emphasizing the need for a standardized framework to model and trace hardware component dependencies, similar to other ecosystems discussed earlier.

Motivated by the success of BOM practices in software and related domains, CISA introduced the Hardware Bill of Materials (HBOM) in September 2023 \cite{cisa:hbomFramework2023} (\ie the rightmost yellow box of the HBOM thread in Fig.~\ref{fig:bom_development}). HBOM aims to trace hardware components in critical infrastructure using hardware-specific data fields, which will be discussed in \S\ref{subsec:bom_structure_analysis}.
HBOM has already demonstrated value in supporting compliance with regulations governing cyber-physical devices (\eg Internet-connected surveillance cameras) used in critical infrastructure \cite{ndia:nationalDefenseAuthorizationAct2019}.

\subsubsection{Software-as-a-Service Applications}
\label{subsubsec:saas_software}

The Software-as-a-Service (SaaS) industry, supported by global cloud infrastructure, has grown rapidly, reaching an estimated market value of USD 273 billion in 2024 with venture capital investments exceeding USD 207 billion \cite{ominus:saasTrendReport2025}. Despite this growth, SaaS applications running in virtualized cloud environments face significant challenges in transparency and trust compared to on-premise solutions, especially for enterprise users. 

Recognizing these concerns, the U.S. government included SaaS transparency in the Executive Order 14028 \cite{federalregister:executiveOrder2021} (\ie the first purple box of the SaaSBOM thread in Fig.~\ref{fig:bom_development}), followed by CISA's release of the first SaaSBOM framework \cite{cisa:saasTransparency2024}, as shown in the second purple box of the SaaSBOM thread in Fig.~\ref{fig:bom_development}. This framework extends existing BOM standards with cloud-specific dependency fields, such as external service names, API endpoints, data classifications, data flows, and authentication requirements, to improve visibility in highly dynamic and virtualized cloud environments.

\begin{figure*}[t]
    \centering
    \captionsetup{skip=2mm}
    \includegraphics[width=0.995\linewidth]{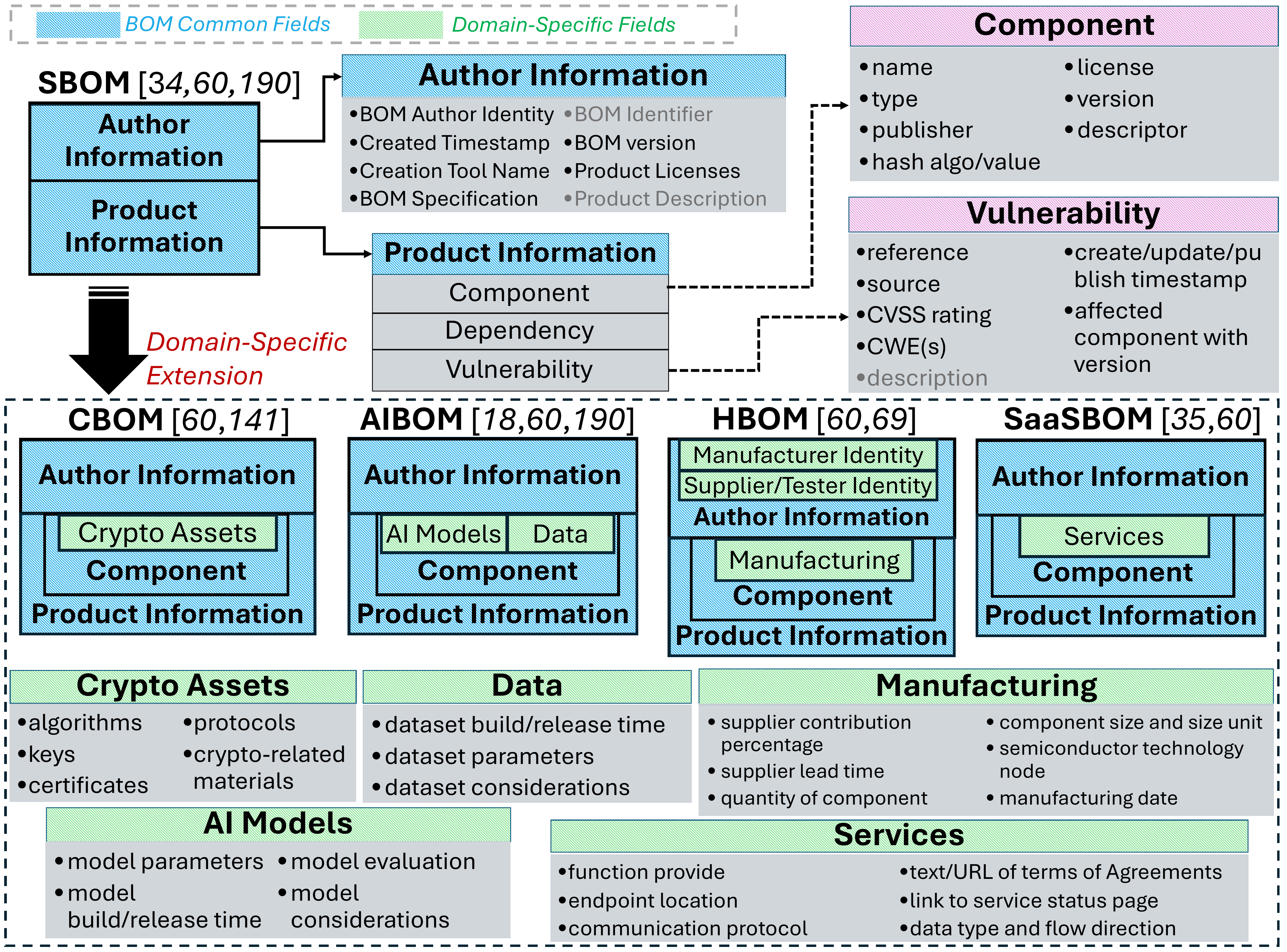}
    \caption{The structures of current BOM frameworks including SBOM, CBOM, AIBOM, HBOM and SaaSBOM, with common data fields in blue color and domain-specific fields in green color.}
    \label{fig:bom_structure}
    \vspace{-7mm}
\end{figure*}

\subsection{Structures of Bill of Materials (BOM) Frameworks}
\label{subsec:bom_structure_analysis}

After introducing the standard BOM frameworks, we now briefly describe their structures as shown in Fig.~\ref{fig:bom_structure}.
In the upper part of the figure, SBOM, the earliest standardized BOM format, consists of two major sections: \myverb{Author Information} and \myverb{Product Information}. The author information (commonly referred to as project metadata) contains required and optional fields, indicated by black and gray texts, respectively. 
For example, the metadata field \myverb{BOM Specification} records the specification version used by the BOM file (\eg SPDX 3.0, CycloneDX 1.7).
The product information section includes three subsections: \myverb{Component}, \myverb{Dependency} and \myverb{Vulnerability}, which list component details, dependency relationships, and vulnerability records (\eg CVE entries) for a software project.

As discussed in \S\ref{subsec:bom_evolution}, SBOM serves as the foundation framework for BOM standards in other digital ecosystems, each extending SBOM with domain-specific fields \cite{github:cycloneDXSpecification2024, spdx:spdxSpecification2024, cisa:sbomMinimumElements2025}. 
The lower part of Fig.~\ref{fig:bom_structure} shows four such frameworks in chronological order. 
CBOM introduces additional fields for crypto assets, including cryptographic algorithms, certificates, encryption keys, and other elements relevant to encrypted systems \cite{cyclonedx:cbomAuthoritativeGuide2024, github:cycloneDXSpecification2024}. AIBOM adds structured descriptions for AI models and their associated training and testing datasets \cite{KBennetLinuxFoundation2024, github:cycloneDXSpecification2024, spdx:spdxSpecification2024}. HBOM incorporates hardware-specific fields in both author information and component sections to cover manufacturer, supplier, and tester identities, as well as detailed hardware manufacturing attributes \cite{cisa:hbomFramework2023, github:cycloneDXSpecification2024}. Finally, SaaSBOM introduces cloud-specific component fields, such as data flows, API types, and endpoint locations, to capture dependencies in dynamic, virtualized cloud environments \cite{cisa:saasTransparency2024, github:cycloneDXSpecification2024}.

\section{BOM Data Generation and Management Practices}
\label{sec:bom_generation_management}

Industry-wide adoption of the Bill of Materials (BOM) to model dependency and transparency in the digital supply chain demands mature tools, platforms, and infrastructure that can support the generation (\S\ref{subsec:bom_tools_repo}), validation (\S\ref{subsec:formal_empirical_evaluation_bom}) and distribution (\S\ref{subsec:bom_sharing_protection}) of BOM data. This section provides a comprehensive survey of the research work in each of these areas.

\subsection{BOM Data Generation Tools and Specialized Repositories}
\label{subsec:bom_tools_repo}
The quality of BOM data, which describes the dependencies of components in a project (\eg software), is highly dependent on the quality of tools used for its generation. Current research has focused on evaluating the capabilities of popular industry tools that produce BOM specifications by analyzing project source codes and/or associated metadata from management systems (\S\ref{subsubsec:evaluating_industry_tools}), developing augmented generation capability (\S\ref{subsubsec:developing_augmented_bom_generators}), and constructing high-quality datasets to benchmark and evaluate BOM generation tools (\S\ref{subsubsec:specialized_bom_data_repo}).

\subsubsection{Evaluating Industry Tools for BOM Data Generation}
\label{subsubsec:evaluating_industry_tools}

Since software projects vary widely across sectors and programming languages, the software industry and the open-source community have developed numerous tools to generate BOM data. Each tool has its own specialization, such as the programming languages and operating systems it supports. To assess how well these tools generate BOM data, researchers have conducted various studies. Most of these studies focus on specific aspects of data quality, such as whether the fields in a BOM are correct \cite{MFRabbiSAC2025, DGarciaSVM2025, MBalliuSP2023, SCofanoTrustCom2024, GBenedettiACNS2025, SYuDSN2024, MFRabbiSAC2024}, complete \cite{STorresAriasSP2023, HVVairagadeNTJ2025, SCofanoTrustCom2024, SYuDSN2024, OSoroceanCONISOFT2024}, and consistent \cite{SYuDSN2024, EODonoghueSCORED2024} between different tools for the same project.

\textbf{Correctness of Generated BOM Data:}
A wide range of industry tools (open-source and proprietary \cite{cyclonedx:ToolCenter2025, spdx:Tools2025}) are available to developers and distributors to generate BOM data for their software products. To ensure that the BOMs data generated by these tools accurately reflect the actual software components of the corresponding products, several research works have benchmarked the similarity between the tool-generated BOMs and the actual software component directories. These benchmarks often rely on package managers and cover popular programming languages, including Rust \cite{MFRabbiSAC2025}, Java \cite{DGarciaSVM2025, MBalliuSP2023, YXiaoNDSS2025}, Python \cite{SCofanoTrustCom2024, GBenedettiACNS2025, SYuDSN2024}, and Javascript \cite{MFRabbiSAC2024}. 
For example, the authors of \cite{MFRabbiSAC2025} evaluated the SBOM data generated by seven open-source tools (\eg cyclonedx-rust-cargo \cite{cyclonedx:cyclonedxRustCargo2020} and spdx-sbom-generator \cite{opensbom:spdxSBOMGenerator2021}) for 50 popular Rust software projects. They measured  how closely the generated SBOMs matched the actual software metadata recorded by the Rust package manager, comparing the component names, versions, and library dependencies. The study found that the generated SBOMs consistently captured fewer components than the ground truth metadata. 
As an example for Java software projects, Garcia \etal \cite{DGarciaSVM2025} benchmarked five popular SBOM generation tools (\eg Syft \cite{anchore:Syft2020}, CycloneDX Generator \cite{cyclonedx:cdxgen2020}) by evaluating the quality of SBOM data produced for a sample Java product in three types of use cases: components that are actually included in the software product, components that are unused but included, and components that are declared but not included. The study identified two major issues in the assessed tools: (1) inconsistencies between declared and unused components in the SBOM and the actual status of the components, and (2) systematic inaccuracies in the studied tools that rely on Java package managers (\eg Maven) as their primary data source.
Beyond these specific examples, most studies in this research stream report accuracy issues in BOM data generated across software products from different programming language communities.

\textbf{Completeness of Generated BOM Data:}
In addition to works that evaluate the accuracy and correctness of BOM data generated by industry tools, there is a body of research that focuses on assessing the completeness of BOM data for key fields of software projects. These fields include software licenses \cite{STorresAriasSP2023}, component metadata \cite{STorresAriasSP2023, HVVairagadeNTJ2025, SCofanoTrustCom2024, SYuDSN2024, YXiaoNDSS2025, RJonesVizSec2023, OSoroceanCONISOFT2024}, and supplier details \cite{STorresAriasSP2023, HVVairagadeNTJ2025}. 
For example, Torres-Arias \etal \cite{STorresAriasSP2023} evaluated all the fields mentioned above in the BOM data from four tools using 1000 Docker images, comparing their results with the NTIA-defined minimum elements for the BOM data. They found that all four tools provided little or no supplier information and one tool (\ie Trivy \cite{aquasecurity:Trivy2019}) included almost no author and timestamp details in its license data. In addition,  none of the tools could produce a full coverage of package IDs and licenses for all components in a project.

\textbf{Consistency Across Generated BOMs:}
Beyond the correctness and completeness of the BOM data, recent research \cite{SYuDSN2024, EODonoghueSCORED2024} has also examined the (in)consistency of the BOM data generated by different open-source tools for the same software projects.
Yu \etal \cite{SYuDSN2024} conducted an empirical study to compare SBOM produced by four popular tools (\eg Microsoft SBOM Tool \cite{microsoft:SBOMTool2022}) for 7,876 open-source repositories on Github. Their analysis revealed various inconsistencies in terms of component dependencies, identified through comparative visualizations and Jaccard similarity measures, as well as duplication in detected software packages (\eg on average, 13.71\% libraries were repeatedly included in SBOM for Python-based projects generated by Microsoft SBOM Tool). Further investigations into the root causes of these inconsistencies highlighted issues such as inconsistent package naming conventions and varying definitions of library and package dependencies introduced during software development.

\subsubsection{Enhancing BOM Data Generation Capabilities}
\label{subsubsec:developing_augmented_bom_generators}

The current tools for generating BOM data for software products help improve transparency, but they still struggle to produce accurate and complete BOM data for large-scale projects. Therefore, an active research area is to develop techniques that improve BOM generation capabilities taking into account challenges such as inaccurate metadata \cite{SNadgowdaWoC2022, ZZhangIC2E2024, NKawaguchiCCNC2024, NSrinivasanarXiv2025, RFosterRWS2021}, missing source code \cite{JJRheeMILCOM2024, VSafronovIoT2024, MBeningerCyCon2024, MBeningerCyCon2024, DPereiraICSPIC2024, LSongarXiv2024}, and rapidly changing ecosystems \cite{EBandaraIWCMC2025, EBandaraCCNC2025, EBandaraMASS2024, HKimICAIIC2024, MAlhanahnaharXiv2024}.

\textbf{Generating BOM Data Beyond Consuming Metadata Files:}
As discussed in \S\ref{subsubsec:evaluating_industry_tools}, existing tools for generating software BOM data often rely on metadata, such as JSON/XML files and package manager records. However, this metadata-only approach overlooks critical runtime software components that are not explicitly listed in metadata repositories. Recognizing this limitation, several research efforts have aimed to extend BOM generation beyond static metadata analysis by incorporating build-time tracing \cite{SNadgowdaWoC2022, ZZhangIC2E2024} and runtime monitoring \cite{NKawaguchiCCNC2024, NSrinivasanarXiv2025, RFosterRWS2021}.
For example, the authors of \cite{SNadgowdaWoC2022} developed \textit{sbom-ctl}, a tool that generates BOM specifications of software components by tracing build-time artifacts, including prebuilt declarations, embedded components, and compiled binaries.
Similarly, Kawaguchi \etal \cite{NKawaguchiCCNC2024} developed the SBOM Container Gateway (SCG), a Docker-based module that identifies missing components from metadata files. SCG achieves this by emulating the runtime behavior of a software project within isolated container environments and comparing the observed components against those declared in package manifests, directories, and version specifications.

\textbf{BOM Data Generation without Access to Source Code:}
In some cases, BOM data must be generated without access to the source code, for example when dealing with legacy binaries, precompiled executables, or proprietary software protected by restrictive licenses. To address these scenarios, the research community has proposed creative BOM generation techniques that analyze binary artifacts directly. These methods often rely on hash-based identification of attributes extracted from software binaries, allowing either (1) exact matching against known library signatures \cite{JJRheeMILCOM2024, VSafronovIoT2024, MBeningerCyCon2024} or (2) probabilistic inference using neural networks to estimate semantic similarity between unknown and known components \cite{MBeningerCyCon2024, DPereiraICSPIC2024, LSongarXiv2024}.
For example, \textit{GrizzlyBay} \cite{JJRheeMILCOM2024} applies locality-sensitive hashing (LSH) to extract attributes such as function blocks, control flows, and abstract machine instructions from legacy binaries, which are then mapped to a reference database of known software component signatures.
To overcome the scalability issues inherent in multi-layer signature matching, \textit{ERS0} \cite{MBeningerCyCon2024} leverages large-scale language models trained in more than 4.3 million software component binaries. This model infers potential library names and version information for a given binary through string-based similarity matching.

\textbf{BOM Data Generation with Augmented Metadata:}
Standard BOM specifications typically identify software components using metadata fields such as component names, versions, and dependencies, as discussed in \S\ref{subsec:bom_structure_analysis}. However, the software research community has noted that these standard metadata fields are often insufficient for capturing domain-specific abstract information, such as data privacy \cite{YXiaoSCORED2024, ZTaoPET2025}, security vulnerabilities \cite{ZSunCCEAI2024}, cryptographic assets \cite{MMoffiearXiv2025}, and software provenance \cite{BSeshadriarXiv2024}.
To address these gaps, several studies have proposed an augmented BOM to support specific use cases. For example, \textit{PoGen} \cite{YXiaoSCORED2024} introduces additional metadata fields related to data privacy, including data storage location, protection methods, data types, and sources. These augmented fields have been applied to multiple popular third-party libraries (\eg Radar SDK), improving the transparency in data flows for users of those libraries.
Another example is \textit{Cryptoscope} \cite{MMoffiearXiv2025} that extends metadata to track cryptographic assets, capturing details such as algorithm variants (\eg AES, SHA256RNG), operation modes, key sizes, and related cryptographic materials (\eg initialization vectors, nonces and salts). This augmented metadata is particularly valuable for the emerging cryptographic software ecosystem.

\textbf{BOM Generation Tools for Specific Ecosystems:}
In addition to augmenting standard BOM specifications with domain-specific metadata, a series of research efforts have developed specialized BOM generation tools to improve the accuracy and completeness of BOM data for software projects within specific programming language ecosystems, including Python \cite{CJiaICSEC2025, GBenedettiACNS2025}, Java \cite{YZhaoISSRE2024}, C/C++ \cite{YChoiICSE2025}, Ansible \cite{ROpdebeeckSANER2025}, and Linux package management systems \cite{TQiuISSTA2025}.
In a recent example, the authors of \cite{ROpdebeeckSANER2025} developed a software composition analysis tool tailored for Ansible plugins. By focusing specifically on Ansible, this tool outperforms generic BOM generation tools, precisely capturing the dependencies of the plugins in the installed packages, the requirements of the platform, and the running versions. 
Another recent tool, \textit{LiPSBOMaker} \cite{TQiuISSTA2025} targets SBOM generation for Linux distribution packages in RPM and Deb formats. Its generation methodology is carefully aligned with the stages of the Linux package distribution lifecycle (from source to release and runtime), enabling it to produce more accurate and complete BOM data for Linux packages than general-purpose tools.

\textbf{AI for BOM Data Generation at Scale:}
As briefly discussed in the preceding paragraphs, several research efforts, such as \textit{ERS0} \cite{MBeningerCyCon2024}, have used AI techniques to automate the generation of BOM data at scale. This approach eliminates the need to manually craft matching rules and conditions for metadata and code analysis; rules can quickly become obsolete in the rapidly evolving software development landscape. A growing body of research work applies AI-based methods to generate BOM metadata \cite{EBandaraIWCMC2025, EBandaraCCNC2025, EBandaraMASS2024} and to construct dependency graphs of software components \cite{HKimICAIIC2024, MAlhanahnaharXiv2024}.
Among various AI approaches, large language models (LLMs) (\eg Meta's Llama model used in \cite{EBandaraCCNC2025, EBandaraIWCMC2025, EBandaraMASS2024}) are becoming increasingly prevalent due to their strong ability to interpret and reason over language-based metadata and source code.
In contrast, feature-based machine learning models, such as regression, decision trees, and neural networks, are typically used to model dependencies among software components \cite{MAlhanahnaharXiv2024}. These models are well-suited for representing relationships in structured graphs with hierarchical dependencies, rather than unstructured textual data, which LLMs process more effectively.

\subsubsection{Specialized BOM Data Repositories}
\label{subsubsec:specialized_bom_data_repo}

In addition to the work that evaluates BOM generation tools and their capabilities, another active area of research focuses on the construction of repositories for generating BOM data. These repositories address community challenges related to the need for ground truth datasets \cite{RKishimotoMSR2025, LSoeiroMSR2025, YGamageMSR2025, STorresAriasSP2023, MMounesanSIN2023, SZacchiroliMSR2022, MJahanshahiMSR2025} and support domain-specific use cases \cite{XRenarXiv2025, WXuACE2023,DKNkodiaFRUCT2022}.

\textbf{Ground Truth Data Repositories:} High-quality ground truth datasets are fundamental for the evaluation and benchmarking of BOM data generation tools. Previous research has produced ground truth repositories of BOM data for a limited number of projects using manual validation \cite{RKishimotoMSR2025}, or constructed large-scale collections through systematic and automated procedures \cite{LSoeiroMSR2025, YGamageMSR2025, STorresAriasSP2023, MMounesanSIN2023, SZacchiroliMSR2022, MJahanshahiMSR2025}. 
Kishimoto \etal \cite{RKishimotoMSR2025} created a BOM dataset for 46 Java software projects through a rigorous manual validation process that corrects inaccurate component dependency references, such as those arising from case-sensitivity issues or inconsistencies inherited from Java package managers (Maven Central).
For large-scale repositories generated through automated logic, the authors of \cite{LSoeiroMSR2025} constructed a collection of 78K SBOM files for projects in the Software Heritage Archive. This effort involved processing more than 94 million source repositories across platforms such as GitHub, GitLab, and PyPI. It included systematic corrections for name inconsistencies, merging repository versions for the same software projects, and handling unknown source file extensions.

\textbf{Repositories for Domain-Specific Use Cases:}
BOM data repositories have also been developed to support domain-specific use cases, including AI models based on neural networks \cite{XRenarXiv2025}, the PyPI software ecosystem \cite{WXuACE2023}, software license analysis \cite{TLiuSANER2024, ZWangCCET2021,QKeICSEC2025}, and software vulnerability analysis \cite{DKNkodiaFRUCT2022}.
As a recent example, Renar \etal \cite{XRenarXiv2025} constructed a Neural Network Bill of Materials (NNBoM) dataset from 55,997 PyTorch repositories. This dataset systematically captures third-party libraries (TPLs), pre-trained models (PTMs), and custom modules used in software projects that employ neural network-based AI models.
Ke \etal \cite{QKeICSEC2025} developed \textit{CLAUSEBENCH}, a repository of license data for projects indexed by open-source software (OSS), enabling fine-grained clause-level license analysis and covering aspects such as distribution rights and obligations. 
Similarly, the authors of \cite{TLiuSANER2024} constructed a BOM repository for 453 software licenses with a specific focus on the SPDX community, incorporating term-level attributes (\eg Copyright), permissions (\eg Commercial Use), and obligations (\eg MUST Include Notice).
As a representative repository for software vulnerability analysis, the work in \cite{DKNkodiaFRUCT2022} built a repository containing XML-based vulnerability data for software projects in the BOM formats adopted by the SPDX and CycloneDX communities. This repository contains essential metadata, including software names, versions, vulnerability types, CVE identifiers, severity levels, attack techniques, and descriptive details.

\subsection{Formal and Empirical Evaluation of BOM Data}
\label{subsec:formal_empirical_evaluation_bom}

After reviewing the research on BOM data generation tools, we now focus on formal and empirical approaches to evaluate the quality of BOM data. These evaluations consider multiple aspects, including completeness (\S\ref{subsubsec:data_completeness}), accuracy (\S\ref{subsubsec:data_accuracy}), compliance (\S\ref{subsubsec:data_compliance}), and effectiveness in supporting downstream tasks (\S\ref{subsubsec:data_effectiveness}).

\subsubsection{Completeness}
\label{subsubsec:data_completeness}Several research efforts have proposed methods to evaluate the completeness of BOM data, focusing on whether the required components and metadata fields are included. These methods formally assess the completeness of the BOM against criteria commonly recognized in the software industry.
The two representative studies \cite{AHalbritterARES2024, SNoceraJSS2025} introduced approaches to assess the completeness of SBOM data with respect to the NTIA's minimum requirements \cite{ntia:sbomMinimumElements2021}, which cover essential fields such as component names, versions, unique identifiers, supplier names, component dependencies, SBOM authors, and timestamps. Halbritter \etal \cite{AHalbritterARES2024} examined software repositories in four programming languages (\ie Python, C, Rust, and Typescript), while Nocera \etal \cite{SNoceraJSS2025} analyzed 186 open-source projects hosted on GitHub. Both studies found that missing fields in the BOM data are a common issue that warrants the community's attention. For example, only 7\% of the SBOM files analyzed in \cite{SNoceraJSS2025} contained all mandatory NTIA fields.

\subsubsection{Accuracy}
\label{subsubsec:data_accuracy}
Ensuring that all compulsory fields are present in a BOM file does not guarantee the correctness of the values within those fields. A group of works has developed verification methods to assess the accuracy of BOM data, focusing on component dependency information \cite{YXiaoNDSS2025, RJonesVizSec2023} and metadata validation \cite{RKishimotoSANER2024}.
Specifically, \textit{JBomAudit}, proposed by \cite{YXiaoNDSS2025}, verifies the accuracy of component dependencies in Java projects by comparing an SBOM file with its corresponding Java binary code. In contrast, \textit{Osmy}, introduced by Kishimoto \etal \cite{RKishimotoSANER2024}, validates the SBOM metadata by comparing the SHA-1 hash values recorded in the SBOM with the checksums computed directly from the source files.

\subsubsection{Compliance with Community Standards}
\label{subsubsec:data_compliance}
As introduced in \S\ref{sec:bom_developments}, two major communities (SPDX and CycloneDX) define standards for BOM data, covering metadata \cite{MScarlatoICOSST2023, QKeICSEC2025, ZWangCCET2021, LSoeiroMSR2025} and schema fields \cite{MScarlatoICOSST2023, LSoeiroMSR2025}.
A representative work by Scarlato \etal \cite{MScarlatoICOSST2023} conducted an empirical evaluation of more than 381,000 open-source packages in the PyPI ecosystem. The authors implemented an automated license-checking process to verify compliance with the SPDX BOM standard and discovered significant quality issues: only 24\% of projects had SPDX-compliant license declarations, 23\% lacked any license metadata, and approximately 2\% experienced HTTP errors during metadata retrieval.

\subsubsection{Effectiveness for Downstream Analysis}
\label{subsubsec:data_effectiveness}
The ultimate measure of BOM data quality is its usefulness in downstream analyses. Existing studies have explored its applications in detecting software vulnerabilities inherited from component dependencies \cite{EODonoghueSANERC2024, GDaliaSSRN2025, EODonoghueSANERC2024, MMounesanSIN2023} and identifying potential license violations in open-source projects that incorporate legally protected components \cite{ZLiuISSRE2024, SXuTSEM2023, SXuISSTA2023, TLiuSANER2024, JWuMSR2024, WXuACE2023}.

\textbf{Vulnerability Detection and Risk Assessment:} BOM data are commonly used to detect and track propagated vulnerabilities in the software supply chain, particularly those arising from vulnerable components such as libraries and functions. Previous research has evaluated BOM effectiveness for individual projects \cite{EODonoghueSANERC2024, GDaliaSSRN2025, EODonoghueSANERC2024}, and entire ecosystems, such as Docker \cite{MMounesanSIN2023}. 
In a comprehensive study, Donoghue \etal \cite{EODonoghueSANERC2024} analyzed 1,151 SBOM files to detect vulnerabilities propagated through third-party libraries. Using static code analysis tools (\ie Trivy and Grype) guided by the component dependencies documented in the BOMs, the authors quantified the occurrences of different vulnerability types and their severity. More than 80\% of the detected vulnerabilities were classified as medium or high severity; issues that would be difficult to identify without SBOM-guided dependency information. 
The authors of \cite{MMounesanSIN2023} focused on the Docker ecosystem, analyzing SBOMs for 12.6K Docker containers encompassing 26K recorded component dependencies on GitHub. Their study revealed systematic security risks, including the widespread use of popular components with severe vulnerabilities and the improper handling of dependency version updates.

\textbf{Analyzing Breaches of Software Licenses:}
Beyond vulnerabilities, the improper use or license violations of software components represent another concern in open-source software. Several studies have investigated the role of BOM data in identifying license breaches \cite{ZLiuISSRE2024, SXuTSEM2023, SXuISSTA2023, TLiuSANER2024, JWuMSR2024, WXuACE2023}.
Xu \etal developed \textit{LiDetector} \cite{SXuTSEM2023}, which extracts software licenses for 1,846 software projects using semi-supervised machine learning models guided by BOM specifications; and \textit{LiResolver} \cite{SXuISSTA2023}, which recommends corrective actions (\eg proper attributions to respective authors) for identified license violations, accounting for hierarchical dependencies captured in BOM files.
Wu \etal \cite{JWuMSR2024} conducted a large-scale analysis in five major package management platforms (\ie Maven, NPM, PyPI, RubyGems, and Cargo), examining license compliance in more than 33 million software projects.
Their findings highlight common issues, such as the failure to acknowledge software licenses (\eg 56.38\% in Maven projects and 15.41\% in PyPI projects), and the use of incompatible licenses (\eg 14.91\% and 13.43\% of Maven and PyPI projects, respectively).

\subsection{Centralized and Decentralized Sharing of BOM Data}
\label{subsec:bom_sharing_protection}

The secure and tamper-resistant sharing of BOM data between producers and consumers has become an increasingly important topic. Current approaches include semi-centralized platforms (\S\ref{subsubsec:bom_sharing_hybrid}), decentralized blockchain-based platforms (\S\ref{subsubsec:bom_sharing_decentralized}), both of which employ cryptographic verification techniques to ensure data integrity.

\subsubsection{Semi-centralized BOM Data Distribution Platforms}
\label{subsubsec:bom_sharing_hybrid}

A large body of work has explored BOM data sharing platforms that combine centralized management with decentralized data distribution architectures \cite{SNadgowdaWoC2022, NKTranEASE2024, JWStokesMILCOM2021}. These hybrid models aim to balance security and efficiency in data distribution with the complexity of management.
One such example is \textit{sbom-ctl} \cite{SNadgowdaWoC2022}, which enforces data-sharing policies and control logic centrally while leveraging decentralized and heterogeneous data distribution mediums, such as Open Container Initiative (OCI) registries, cloud object storage (COS), and distributed file systems, for efficiency. In this design, data storage and sharing are decentralized, whereas access privileges and policy enforcement remain under a centralized control plane to maintain security.
Similarly, the \textit{SCM2} system \cite{NKTranEASE2024} supports the sharing of SBOM data within an organization by managing SBOM metadata through a centralized domain that handles authentication and notarization. Once metadata verification is complete, the actual BOM content is distributed in a decentralized manner, providing security and scalability.

\subsubsection{Decentralized Blockchain-Based Approaches}
\label{subsubsec:bom_sharing_decentralized}

Fully decentralized BOM data sharing platforms, operating independently of central hosting or administrative authority, have been developed, particularly for the open-source community. Many of these platforms leverage blockchain technologies for their privacy-preserving, auditable, and tamper-evident properties \cite{BXiaSVM2024, XZhouJSEP2024,YLiuICB2024, RBoharaSEW2023, ASongICACT2025, DEHyeonAsiaJCIS2023}.
For example, Bohara \etal \cite{RBoharaSEW2023} proposed a decentralized SBOM sharing framework built on the InterPlanetary File System (IPFS). Their design employs addressable off-chain storage indexed by immutable on-chain SBOM metadata, which is protected by digital signatures and transaction hashes at the granularity of software versions.
Similarly, the authors of \cite{DEHyeonAsiaJCIS2023} developed a decentralized SBOM data sharing for the IoT ecosystem. Their system stores SBOM data, IoT firmware images, and manifest files in IPFS, while maintaining metadata (\ie index) on private or consortium blockchain networks.
Beyond decentralized BOM sharing platforms, several studies have focused on verification techniques used within these systems, such as digital signatures \cite{TRSchorlemmerSP2025, JWStokesMILCOM2021} and hash-based verification functions \cite{COzkanarXiv2025, OStengeleICBC2025}, to ensure data authenticity and integrity.

\section{Applying BOM Data for Transparency and Dependency Analysis}
\label{sec:bom_application}

BOM data provide detailed information on the components used in a product and their associated metadata. This makes BOM data a valuable resource for security researchers in several ways: to model software component dependencies (\S\ref{subsec:bom_modelling_components}), to validate component compliance and assess operational risks (\S\ref{subsec:bom_compliance_validation}), and to trace the propagation of vulnerabilities in proactive impact assessment and forensic analysis (\S\ref{subsec:bom_vulnerability_analysis}).

\subsection{Modeling Component Inter-relationship and Dependency}
\label{subsec:bom_modelling_components}
As intended, BOM data have been widely used by researchers to construct comprehensive inventories for digital projects: encompassing not only software and hardware systems, but also emerging areas such as AI models and cryptographic assets (\S\ref{subsubsec:bom_modelling_components}). These inventories form the foundation for security analysts to model inter-component dependencies and identify propagated vulnerabilities (\S\ref{subsubsec:bom_modelling_dependency}).

\subsubsection{Constructing Component Inventory of Digital Products}
\label{subsubsec:bom_modelling_components}
The most direct application of BOM data is to provide users with a comprehensive view of all components used in digital products. This approach is relatively established for more mature BOM types such as SBOM and HBOM, but emerging domains, such as AI models \cite{IBarclayCCPE2023, AMakrisCSR2025, QLuSoftware2024, TStalnakerICSEC2025, PRadanlievJDMS2024} and cryptographic assets \cite{VNethenEICC2024, MBeningerCyCon2024, BHessETSI2024, HWLimNCS2024}, are still in the early stages of adoption and continue to drive ongoing research efforts.

\textbf{AI/ML Models and Datasets:}
AI models and their training and evaluation datasets have become increasingly critical in various industry sectors. Ensuring the reliable use of AI systems requires the BOM community to accurately describe the unique dependencies inherent in AI models to assess potential biases and vulnerabilities. Therefore, recent research has explored BOM applications tailored for AI models that capture their unique dependency characteristics, such as training data provenance \cite{IBarclayCCPE2023, AMakrisCSR2025, QLuSoftware2024, TStalnakerICSEC2025, PRadanlievJDMS2024}, model architecture specifications \cite{TStalnakerICSEC2025, PRadanlievJDMS2024} and algorithmic mechanisms \cite{PRadanlievJDMS2024}. 
As an early example, Barclay \etal \cite{IBarclayCCPE2023} introduced a BOM framework that leverages self-sovereign identity protocols and cryptographically signed credentials to record metadata for AI models and their training datasets. This includes identifiers (\eg model name and version, dataset name and roles), download locations (\eg URLs), contributor information (\eg name, email, role such as ``programmer''), and ethical sourcing status. 
Building on this, Radanliev \etal \cite{PRadanlievJDMS2024} proposed the AIBOM (Artificial Intelligence Bill of Materials) schema, which formally organizes AI-specific metadata into a standardized JSON structure.

\textbf{Cryptographic Assets:}
Cryptographic assets are also emerging as digital products that require a standardized BOM specification to capture their inherent complexity. Such complexity includes algorithm types and key lengths, security parameters and certification status, implementation dependencies, asset relationships, and considerations for post-quantum cryptography (PQC) migration and lifecycle management \cite{VNethenEICC2024, BHessETSI2024, HWLimNCS2024}.
In 2024, Hess \etal \cite{BHessETSI2024} introduced the first cryptographic BOM (CBOM), a metadata document to describe cryptographic assets and their dependencies within software systems. Integrated into the CycloneDX 1.6 object model, the CBOM includes metadata on algorithms, protocols, certificates, and cryptographic material (\ie security levels, certification status, and implementation environments). A CBOM document also captures relationships between assets using the \textit{dependsOn} field, along with their proprieties (\eg supported applications and TLS protocol versions).
Building on this foundation, subsequent studies \cite{VNethenEICC2024,HWLimNCS2024} have used CBOM to create comprehensive inventories of cryptographic assets and analyze their resilience against quantum-based decryption threats.

\subsubsection{Dependency Modeling}
\label{subsubsec:bom_modelling_dependency}
With BOMs providing systematic inventories of assets and their components, researchers have started developing methods to model component dependencies directly from BOM data. Recent work has used BOM-derived component metadata to construct dependency graphs \cite{YGamageMSR2025, SSpethICSAC2025} and to develop (interactive) visualization frameworks capable of representing complex dependency structures that exceed the expressiveness of simple data models \cite{BAKellerINL2024, ROberhauserBMSD2025}.
For example, Speth \etal \cite{SSpethICSAC2025} model inter-component relationships in Kubernetes environments by extracting library-level dependency metadata from SBOMs, enabling the construction of comprehensive dependency graphs that reveal how container vulnerabilities may propagate. Oberhauser \etal \cite{ROberhauserBMSD2025} developed an interactive visualization tool that converts SBOM data into 3D, directionally linked graphs, helping security analysts more easily interpret complex and hierarchical dependency structures.

\subsection{Compliance Validation and Operational Risk Assessment}
\label{subsec:bom_compliance_validation}

BOM data, which describe components included in digital projects together with metadata on their associated licenses, has become an important resource for researchers. It is used to evaluate the license compliance of the included components (\S\ref{subsubsec:bom_compliance_licensing}); assess whether a project adheres to government regulations (\S\ref{subsubsec:bom_compliance_regulatory}); verify the integrity and provenance of the dependent components throughout the product lifecycle (\S\ref{subsubsec:bom_compliace_integrity}); and analyze security vulnerabilities that may propagate through those dependencies (\S\ref{subsubsec:bom_compliance_risk_assessment}).

\subsubsection{Licensing and Intellectual Property Compliance}
\label{subsubsec:bom_compliance_licensing}

With the availability of BOM data for digital products, researchers have developed methods to verify license compliance in open-source and commercial projects \cite{JNMazonBigData2021, LColonnaCLSR2025, TLeeMetaCom2023}, as well as for emerging digital products such as AI models \cite{QLuCAIN2024}.

\textbf{License Compliance for Open-source and Commercial Projects:}
Validating software license by tracking components through BOM data is becoming a mature industry practice in many sectors. However, a new research challenge is emerging as nontraditional software components (\eg datasets) start to play an important role in licensing workflows. For example, Mazon \etal \cite{JNMazonBigData2021} proposed a compliance assessment framework for open data reuse that extends the DCAT (Data Catalog Vocabulary) standard with SPDX (Software Package Data Exchange) fields. This extension allows each dataset incorporated into a software project to be represented as a traceable component with formal licensing metadata and provenance information.
BOM data has also been used for intellectual property (IP) management purposes \cite{LColonnaCLSR2025, TLeeMetaCom2023}, as it provides structured documentation that enables researchers and practitioners to construct detailed dependency maps of licensed assets for liability attribution.
For example, the authors of \cite{LColonnaCLSR2025} use BOM data to examine downstream manufacturers' compliance with licensing obligations under the EU Cyber Resilience Act (CRA) \cite{europa:cyberResilienceAct2024} and the Product Liability Directive (PLD) \cite{europa:productLiabilityDirective2024}.
Using the licensing fields in SBOM data, Lee \etal \cite{TLeeMetaCom2023} developed a blockchain-based framework to track copyright and license-agreement histories in outsourced software development. Their framework enables automated recording of the software composition and associated contractual metadata, including copyright transfers, license terms, and payment records between vendors and clients, ensuring end-to-end traceability and version control throughout the development lifecycle.

\textbf{AI/Data Component Licensing:}
Emerging AI and machine learning systems broaden the scope of license compliance beyond traditional software components, introducing new considerations around training data licensing, model ownership, and the rights associated with derivative model development. To address these AI-specific compliance requirements, recent work has used the newly proposed AIBOM standard \cite{QLuCAIN2024} to perform license checks across the AI supply chain. AIBOM enables structured documentation and verification for externally sourced foundational models (FMs) (\eg whether open-source versus commercial, general-purpose versus domain-specific), as well as for fine-tuned models characterized by their parameter sets, and distilled variants ranging from lightweight to heavyweight. It also incorporates verifiable Responsible AI (RAI) credentials and governance metadata, such as approved and/or prohibited use cases, model versioning, and input/output specifications.

\subsubsection{Regulatory and Policy Compliance}
\label{subsubsec:bom_compliance_regulatory}
BOM frameworks are increasingly becoming mandatory instruments for meeting sector-specific regulatory requirements. The work in \cite{ARNygardPCI2024, SCarmodyNPJDM2021} highlighted how BOM-based approaches support compliance with recent EU cybersecurity legislation, particularly the NIS2 Directive, the Cyber Resilience Act (CRA), and the Network Code on Cybersecurity for the Electricity Sector (NCCES). Collectively, these regulations require critical infrastructure operators to implement comprehensive digital supply chain risk-management practices, including continuous assessment of supplier relationships, verification of component security, and digital asset life cycle tracking. To operationalize these regulatory obligations, researchers have used the SBOM and HBOM frameworks as foundational tools for structured documentation throughout the procurement, deployment, and maintenance phases. These BOM-driven workflows enable organizations to systematically meet mandated transparency and control measures, such as vulnerability disclosure, supplier security assessment, and verification of digital component provenance.

\subsubsection{Integrity and Provenance}
\label{subsubsec:bom_compliace_integrity}

Beyond license compliance, BOM data has been used in research to verify the integrity of software components at two critical stages of the software lifecycle: build-time integrity assurance \cite{KLewAS2024, DGamaLADC2024} and runtime provenance verification \cite{YShimamotoNDSS2025, ASharmaarXiv2024, NKawaguchiCCNC2024}.

\textbf{Build-Time Integrity Assurance:}
The build-time stage, as the initial stage of the product assembly process, establishes the foundation for documenting the software components and their metadata to ensure integrity. To create trustworthy baselines, researchers have utilized reproducible build technologies and cryptographic attestations \cite{KLewAS2024, DGamaLADC2024} to produce verifiable BOM data, which serves as authoritative evidence for integrity validation.
For example, Lew \etal \cite{KLewAS2024} proposed a distributed software build assurance framework to verify the integrity of software artifacts. Their approach constructs Merkle trees from software artifacts, including source code files, reproducible build outputs, and SBOM documentation, and generates cryptographic proofs using hash functions. By storing these proofs on a blockchain, this framework enables users to independently verify the integrity of the software by comparing locally generated proofs with references on the blockchain, eliminating the need for real-time interaction with centralized servers.

\textbf{Runtime Provenance Enforcement:}
After ensuring build-time integrity, maintaining the operational provenance of software components becomes important. Researchers have developed runtime validation mechanisms that use BOM data to enforce execution policies, including access control-based enforcement \cite{YShimamotoNDSS2025, ASharmaarXiv2024} and dynamic behavior monitoring \cite{NKawaguchiCCNC2024}.
For example, the authors \cite{ASharmaarXiv2024} introduced \textit{SBOM.EXE}, a runtime integrity-enforcement system for Java applications that leverages a build-time generated Bill of Materials Index (BOMI) as a whitelist for authorized class loading. The system analyzes application dependencies, standard libraries, and dynamically generated classes to construct a BOMI that enumerates permitted bytecode artifacts. At runtime, a watchdog component intercepts class-loading events, validates each class against the BOMI to ensure provenance, and terminates execution when detecting classes that are unauthorized or tampered with.
Another work by Kawaguchi \etal \cite{NKawaguchiCCNC2024} proposed \textit{C3DRS}, a compliance-aware container deployment framework that enforces risk-based access control through SBOM-based dynamic behavior monitoring. The framework executes container images in a controlled environment, observing file access and execution behaviors to generate SBOMs that reflect components loaded during installation. These SBOMs also serve as baselines for runtime provenance verification by continuously tracking system calls corresponding to the components that are actively loaded during execution.

\subsubsection{Risk Assessment}
\label{subsubsec:bom_compliance_risk_assessment}

In addition to regulatory compliance requirements, effective decision-making by digital product users also relies on systematic methods for operational risk assessment that leverage BOM data. Recent research has explored governance-driven risk scoring for component selection \cite{ZWuACL2025}, transparency visualization for procurement decisions \cite{PJCavenSSRN2024}, and dynamic attestation mechanisms for runtime compliance enforcement \cite{DGamaLADC2024}.
The authors \cite{ZWuACL2025} proposed \textit{LIBVULNWATCH}, an LLM-driven multi-agent framework to evaluate open-source AI libraries in five governance-oriented dimensions: licensing, security, maintenance, dependency management and regulatory compliance. Within the dependency management dimension, the system evaluates the completeness of SBOM data as an indicator of the trustworthiness of components and the transparency of overall supply chain.
The work by Gama \etal \cite{DGamaLADC2024} proposed a runtime compliance enforcement framework that models vulnerability posture within a Zero-Trust admission control policy. This approach uses SBOM as baselines for labeling risky runtime components.

\subsection{Traceability for Vulnerability and Root-Cause Analysis}
\label{subsec:bom_vulnerability_analysis}
BOM data enables two complementary forms of traceability in digital supply chains ecosystem. First, \textbf{vulnerability-to-impact} analysis (\S\ref{subsubsec:bom_vulnerablity_trace}) helps analysts determine how newly disclosed vulnerabilities may affect their products through dependent components. Second, \textbf{root-cause} analysis (\S\ref{subsubsec:bom_challenges_traceability}) uses BOM data to identify the specific components responsible for security incidents in systems with complex dependency structures.

\subsubsection{Vulnerability-to-Impact Traceability}
\label{subsubsec:bom_vulnerablity_trace}
Public databases such as CVE, NVD, and OSV record known product vulnerabilities, but security teams must still determine (often under tight time constraints) whether these vulnerabilities affect their own systems. To support this need, researchers have developed traceability methods that combine BOM data with vulnerability databases to automatically identify impacted components and products. Existing work spans different stages of a product lifecycle, including pre-deployment analysis \cite{OKagizmandereIJISS2024, AMoodutoINCITEST2023, RBoharaSEW2023, HXiaarXiv2025, YDingICFTIC2024, ATacchellaTCPS2025, SKwonICEA2023, DEHyeonAsiaJCIS2023, RKishimotoSANER2024, YZhaoISSRE2024, VAshiwalEuroSPW2024}, runtime detection \cite{AFasanoACSAC2024, ACrawfordSCORED2023, ECornelissenarXiv2025, DGamaLADC2024} and full lifecycle monitoring \cite{SSteinCSBJ2025, OVZaritskyiSMARTCOMP2025, LSBeeviICVADV2025, WWuICICN2023, MBeningerCyCon2024, MEArangurenSSRN2025, BVanRAMS2024, VSafronovIoT2024, GDaliaSSRN2025, MMoffiearXiv2025}.

\textbf{Tracing Vulnerability in the Pre-Deployment Stage:}
Identifying vulnerabilities in a software project before release is essential, especially when the project depends on third-party components. Previous work has developed BOM-based methods that help developers detect these issues early, either by integrating them into CI/CD pipelines \cite{OKagizmandereIJISS2024, AMoodutoINCITEST2023, RBoharaSEW2023}, running them as standalone automated workflows \cite{HXiaarXiv2025, YDingICFTIC2024, VAshiwalEuroSPW2024}, or tailoring them to specific domains \cite{ATacchellaTCPS2025, SKwonICEA2023, DEHyeonAsiaJCIS2023, RKishimotoSANER2024, YZhaoISSRE2024}.
As an example of CI/CD-integrated methods, Kagizmandere \etal \cite{OKagizmandereIJISS2024} proposed a three-stage method that (1) generates SPDX-compliant manifests using Microsoft's SBOM tool, (2) matches the components in SBOM against the MITRE-CVE database, and (3) uses the Exploit Prediction Scoring System (EPSS) scores to estimate the probability of exploitation and identify mitigation options.
As an example of automated detection workflows, Ashiwal \etal \cite{VAshiwalEuroSPW2024} built a workflow that combines SBOM data with CSAF Vulnerability Exploitability eXchange (VEX) alerts. Their system uses large language models to extract vulnerability information from unstructured sources (\eg supplier emails, blogs, and newsletters) and converts it into machine-readable CSAF VEX format. These records are then mapped to the components listed in an SBOM to determine which parts of a project are affected.
Several works design tracing systems for specific ecosystems. For example, in the Docker ecosystem, \textit{VDIRS} \cite{SKwonICEA2023} automates vulnerability detection and mitigation for Docker images using SBOM data. It scans components against external databases and labels the findings with CVSS severity scores, allowing proactive vulnerability assessment in containerized environments.

\textbf{Tracing Vulnerability during Software Runtime:}
After the software is deployed, organizations must continuously monitor for newly disclosed vulnerabilities and detect unauthorized or unexpected file or component behaviors. Several research efforts use SBOM data, combined with kernel-level runtime signals, to trace vulnerabilities in deployed systems \cite{AFasanoACSAC2024, ACrawfordSCORED2023, ECornelissenarXiv2025}.
For example, the authors of \cite{ACrawfordSCORED2023} proposed a runtime enforcement mechanism that detects the execution of unauthorized or vulnerable components inside Docker containers by referencing each container's SBOM. Their system (1) identifies the expected software components from the SBOM, (2) collects Linux kernel metrics (\ie IMA and eBPF) that capture file access behavior at runtime, and (3) flags abnormal or suspicious activity that deviates from the SBOM. Another example is \textit{NodeShield} \cite{ECornelissenarXiv2025}, a framework to trace vulnerability at runtime in Node.js applications. NodeShield continuously monitors the modules listed in an application's SBOM by tracking imported modules and the use of privileged APIs that access files, network interfaces, and other system resources at runtime.

\textbf{Vulnerability Tracing for the Entire Software Lifecycle:}
Beyond pre-deployment or runtime stages, some systems are designed to trace vulnerabilities across the entire software lifecycle, covering a wide range of ecosystems, including safety-critical systems \cite{SSteinCSBJ2025, BVanRAMS2024}, IoT platforms \cite{OVZaritskyiSMARTCOMP2025, LSBeeviICVADV2025, WWuICICN2023}, and embedded firmware \cite{MBeningerCyCon2024, VSafronovIoT2024, GDaliaSSRN2025, MEArangurenSSRN2025, MMoffiearXiv2025}.
For example, Van \etal \cite{BVanRAMS2024} developed a dynamic vulnerability monitoring pipeline for cyber-physical IoT systems. The pipeline uses SBOM data to identify dependencies on software packages with known vulnerabilities from sources such as NVD and VulDB. This information establishes a baseline for continuous runtime monitoring. Similarly, Beevi \etal \cite{LSBeeviICVADV2025} focus on IoT-based smart delivery systems. They leverage SBOM data to build a comprehensive inventory of software components in the IoT application and use it to assess vulnerabilities and trace potential issues throughout the system's operational lifecycle.

\subsubsection{Tracing Vulnerability for Root-Cause Analysis}
\label{subsubsec:bom_challenges_traceability}

BOM data are also valuable for post-incident analysis, where investigators must determine which component triggered or enabled a security failure. Because BOMs capture detailed dependency relationships, they help analysts trace an incident back to the underlying vulnerable or misconfigured component. 
For example, Okafor \etal \cite{COkaforSCORED2022} developed a forensic analysis approach that uses BOM metadata to assess the transparency, validity, and dependency structure of software supply chains. Their system catalogs component metadata and reconstructs logical links between components during deployment. This allows downstream users to identify which components are affected during an incident and isolate the root cause more efficiently.

\section{Limitations, Refinements, Extensions and Novel Conceptual Models}
\label{sec:bom_extension}
Although BOM frameworks have become essential tools for digital supply chain management, real-world deployments have revealed several structural, operational, and domain-specific limitations. Prior studies have identified gaps (\S\ref{subsec:bom_limitation_study}) in interoperability, semantic expressiveness, security mechanisms, and coverage for emerging digital ecosystems. To address these challenges, researchers have proposed refinements and extensions to existing BOM standards (\S\ref{subsec:bom_limitation_refinement}) and introduced new BOM models tailored to domains such as data ecosystems, AI systems, and dynamic deployment environments (\S\ref{subsec:bom_limitation_new_models}).

\subsection{Limitations of BOM Frameworks}
\label{subsec:bom_limitation_study}
BOM frameworks have received significant attention from government agencies, standards bodies, and researchers. 
As their adoption grows, several practical limitations have been identified across real-world deployments. Prior studies highlight issues including: incompatibilities between BOM standards developed by different communities (\S\ref{subsubsec:bom_limitation_study_standard}), structural and semantic gaps in how BOMs represent dependencies (\S\ref{subsubsec:bom_limitation_study_structural}), insufficient cryptographic protections for secure sharing and verification (\S\ref{subsubsec:bom_limitation_study_security}), limited support for emerging ecosystems and component types (\S\ref{subsubsec:bom_limitation_study_domain}), gaps that hinder security use cases (\S\ref{subsubsec:bom_limitation_study_compliance}), and organizational barriers barriers that slow down the industry adoption (\S\ref{subsubsec:bom_limitation_study_organization}).

\subsubsection{Incompatibility of BOM Frameworks by Different Communities}
\label{subsubsec:bom_limitation_study_standard}

Researchers have identified interoperability as one of the most persistent barriers to large-scale BOM adoption.  Multiple standards, such as SPDX, SWID, and CycloneDX, produce BOM data in different formats, leading to incompatibilities that complicate analysis and automation. \cite{GDaliaARES2024, BKloegASIACCS2024, NZahanSP2023,DButtnerDTIC2022, BKloegASIACCS2024}.
Zahan \etal \cite{NZahanSP2023} analyzed 200 industry reports and found that inconsistent identifiers for software components and versions across these standards are a major obstacle. 
When projects involve hundreds or thousands of dependencies, mismatched identifiers make cross-system comparisons, vulnerability tracking, and dependency analysis unnecessarily difficult.  
Similarly, Buttner \etal \cite{DButtnerDTIC2022} highlighted semantic inconsistencies between SPDX and CycloneDX, showing that the same component can be represented differently depending on the standard used. In general, the lack of alignment between BOM standards has become a pressing issue, underscoring the need for unified identifiers, harmonized schemas, and improved cross-standard mappings to support scalable and reliable supply-chain analysis.

\subsubsection{Structural and Semantic Deficiencies}
\label{subsubsec:bom_limitation_study_structural}

Current BOM frameworks also face structural and semantic gaps that limit their ability to fully capture component dependencies. The issues generally stem from two sources:  incomplete metadata coverage \cite{BXiaICSE2023, TStalnakerICSE2024, MAEremeevACCS2023} and insufficient support for multilayer or hidden dependencies \cite{BXiaICSE2023}. 
Several studies have shown that today’s BOM specifications (\eg SPDX, CycloneDX, SWID) omit the necessary metadata needed for risk assessment.
For example, Eremeev \etal \cite{MAEremeevACCS2023} found that SBOMs often lack fields such as code quality metrics, developer activity patterns, and community governance indicators that are important for assessing the risk of using open-source components.
Other works highlight limitations in representing deeper dependency chains. 
Xia \etal\cite{BXiaICSE2023} showed that BOMs generated under SPDX and CycloneDX typically capture only direct (single-layer) dependencies, leaving many transitive or nested dependencies undocumented. This creates blind spots where vulnerabilities can remain undetected in complex dependency graphs.

\subsubsection{Insufficient Cryptographic Protections}
\label{subsubsec:bom_limitation_study_security}

The integrity of BOM data depends on strong cryptographic protections during storage and sharing. However, current BOM standards still exhibit several weaknesses. Prior studies highlight three main issues: (1) use of outdated hash functions that no longer provide reliable tamper resistance \cite{LJCampSemanticScholar2021}, (2) limited resilience against post-quantum attacks, leaving BOM signatures and integrity checks vulnerable to future cryptographic breaks \cite{LJCampSemanticScholar2021, AGeremewIJEST2024}, and (3) lack of independent or third-party verification mechanisms, making it difficult to validate the authenticity and provenance of shared BOM files \cite{LJCampSemanticScholar2021}.

\subsubsection{Inadequate Coverage of Emerging Ecosystems}
\label{subsubsec:bom_limitation_study_domain}

Existing BOM standards were designed primarily for traditional software and, therefore, lack support for newer categories of digital products. Recent studies highlight gaps in the representation of AI/ML systems \cite{KSinghCSNet2024, MDuanarXiv2024, TBiTSEM2024, TStalnakerICSE2024} and cyber-physical systems \cite{TStalnakerICSE2024}.
For example, Singh \etal \cite{KSinghCSNet2024} showed that current SBOM formats cannot express critical AI metadata, such as training data provenance and versioning, model architectures, hyperparameters and dependencies across data-processing and inference pipeline.

\subsubsection{Insufficient Supports for Risk Assessment and Compliance Validation}
\label{subsubsec:bom_limitation_study_compliance}

Researchers have also highlighted limitations in the use of BOM specifications for cybersecurity risk assessment and compliance validation.
Although BOM data provide visibility into component dependencies, it lacks security-focused metadata specifically needed for deeper analysis \cite{GAWeaverSP2025, BKloegASIACCS2024, MAEremeevACCS2023, DButtnerDTIC2022}.
For example, Weaver \etal \cite{GAWeaverSP2025} examined  several major security incidents and identified three key gaps in today's BOM frameworks: (1) they cannot  represent how risks propagate through transitive third-party relationships, (2) they provide limited visibility into implicit dependencies and privilege boundaries that shape the attack surface, and (3) they lack support for modeling threats across the full system lifecycle.

\subsubsection{Barriers for Integration BOM into Enterprise Operational Processes}
\label{subsubsec:bom_limitation_study_organization}

Researchers have also examined the practical obstacles that enterprises face when incorporating BOM data into daily operational flows.
Geremew \etal \cite{AGeremewIJEST2024} identified several challenges in maintaining CBOM for cryptographic inventory management, including organizational inertia, limited expertise at both the leadership and technical levels, and the high cost of maintenance and updating BOM-related systems.
More broadly, Dawoud \etal \cite{ADawoudEuroSPW2024} analyzed barriers that hinder the adoption of routine BOM in enterprise security operations, such as governance deficiencies, fragmented tooling ecosystems with incompatible functionalities, and coordination difficulties between business units.

\subsection{Refining and Extending BOM Frameworks}
\label{subsec:bom_limitation_refinement}

After identifying the limitations of BOM frameworks, researchers have proposed a range of refinements and extensions to address them. These efforts focus on improving the granularity and completeness of dependency descriptions (\S\ref{subsubsec:bom_refinement_granularity_enhancement}), integrating security-related metadata for product components to facilitate vulnerability analysis (\S\ref{subsubsec:bom_refinement_security_integration}), improving support for project-level risk assessment (\S\ref{subsubsec:bom_refinement_risk_assessment}), expanding coverage throughout all phases of the software lifecycle (\S\ref{subsubsec:bom_refinement_structural_refinement}), and adapting BOM specifications for emerging digital ecosystems (\S\ref{subsubsec:bom_refinement_architecture_adaptation}).

\subsubsection{Enhancing Granularity and Dependency Modeling Capability}
\label{subsubsec:bom_refinement_granularity_enhancement}

Current BOM standards often lack the granularity needed to capture fine-grained or deeply layered dependencies. To address this limitation, previous work has introduced refinements that improve the depth and precision of dependency modeling. Some approaches improve visibility across the lifecycle by continuously monitoring component relationships throughout development, deployment, and operation \cite{TQiuISSTA2025, BSeshadriarXiv2024, HOkhraviSP2025}. Other efforts focus on enriching the BOM metadata with finer details, such as the specific functions and data structures used within each component, enabling more precise vulnerability and impact analysis \cite{YChoiICSE2025, YZhaoISSRE2024, XRenarXiv2025}.

\textbf{Dependency Modeling throughout Project Lifecycle:}
Component dependencies can change across different stages of a digital product, from development to compilation and runtime. Traditional BOMs mainly capture dependencies declared in source code and repositories, which limits their ability to reflect the components actually used throughout the full lifecycle \cite{TQiuISSTA2025, BSeshadriarXiv2024}.
As a recent example, Qiu \etal \cite{TQiuISSTA2025} proposed a multi-stage model for Linux distribution packages that tracks dependencies across source, release, and runtime phases.
Similarly, Seshadri \etal \cite{BSeshadriarXiv2024} introduced a layer-provenance model that extends BOMs to capture build-time dependencies often missed by standard frameworks. Their approach constructs dependency graphs linking source artifacts, compiled outputs, and external SPDX documents to reveal hidden dependencies during the build process.

\textbf{Fine-grained Dependencies in Functions and Data Structures:}
Current BOM frameworks generally track dependencies at the component level (\eg software modules or libraries) but do not provide fine-grained visibility into which functions or data structures are actually used. To address this, several works have extended BOMs to capture detailed dependency information \cite{YChoiICSE2025, YZhaoISSRE2024,XRenarXiv2025}. 
For example, Zhao \etal \cite{YZhaoISSRE2024} developed \textit{CovSBOM}, which traces function call chains in third-party libraries, enabling precise vulnerability assessment at the function level.
Ren \etal \cite{XRenarXiv2025} developed \textit{NNBoM}, an extension for machine-learning projects that captures metadata of neural network structures, such as invocation patterns of pre-trained models (\eg HuggingFace or PyTorch Hub) and their application domains (\eg natural language process or computer vision).

\subsubsection{Integrating Data Fields for Vulnerability Analysis}
\label{subsubsec:bom_refinement_security_integration}

Current BOM frameworks focus on capturing component dependencies, but provide limited support for vulnerability analysis, such as vulnerability metadata (\eg patch status) and cryptographic verification details for each component. To address this gap, researchers and security communities have developed methods to integrate this security-related information into BOMs \cite{DFucciFSEC2025, HOkhraviSP2025,JJRheeMILCOM2024, SMeiklejohnarXiv2025}.

\textbf{Metadata for Vulnerability Contexts:}
Vulnerability information for software components is typically maintained in databases such as CVE. To better support vulnerability analysis, researchers have proposed extending BOM frameworks with metadata linking components to relevant entries in these databases.
For example, Fucci \etal \cite{DFucciFSEC2025} extended SBOMs to include fields from the Vulnerability Exploitability eXchange (VEX) database, capturing vulnerability indexes, status (\ie affected, not affected, fixed, or under investigation), patch information, and mitigation timelines. This provides a more comprehensive view of the vulnerability landscape for a software project.

\textbf{Cryptographic Signatures for Component Verification:}
Current BOM frameworks typically include basic hash values to allow consumers to verify the integrity of a software component (\eg source code) against its BOM entry. However, simple hashes can be bypassed by malicious actors who can alter components without being detected \cite{SMeiklejohnarXiv2025}. To improve verification, researchers have proposed cryptographic enhancements. 
For example, Meiklejohnar \etal \cite{JJRheeMILCOM2024} introduced locality-sensitive hash signatures as a BOM field, which better preserve the binary structure of a component and provide stronger integrity checks than standard hashes.

\subsubsection{Structural Revision for Risk Assessment}
\label{subsubsec:bom_refinement_risk_assessment}
Beyond improving individual BOM fields, some research has focused on structural revisions of BOM frameworks to support comprehensive risk assessments. These revisions aim to meet regulatory requirements\cite{PGustavssonSSRN2025, NLopezRWS2022, OStengeleICBC2025} and improve trustworthiness across digital supply chains \cite{ABasuIFIPTM2023}.

\textbf{Regulatory Compliance:}
Driven by new regional and international regulations, such as the EU Cyber Resilience Act, researchers have proposed structural revisions to BOM frameworks to better support regulatory compliance. Those revisions add fields for supplier metadata, component compliance ratings, regulatory exposure indicators, and transitive risk models \cite{PGustavssonSSRN2025, NLopezRWS2022, OStengeleICBC2025}, complementing the dependency information already captured by standard BOMs. 
As a specific example, Lopez \cite{NLopezRWS2022} enhanced SBOM and HBOM frameworks to improve supply chain risk management. The revision includes corporate information for each component, such as financial status and organizational relationships, in addition to technical details like hardware specifications, software versions, and functions. This helps regulatory authorities identify responsible organizations in the event of cybersecurity incidents.

\textbf{Trustability of Digital Supply Chains:}
Industry stakeholders increasingly require ways to assess the trustworthiness of digital supply chains, which has motivated revisions to the BOM data schema. Recent work proposes adding reputation-related fields to BOMs \cite{OStengeleICBC2025, ABasuIFIPTM2023}, such as code quality, community governance practices, contributor reputation, and deployment requirements.
For example, Basu \etal \cite{ABasuIFIPTM2023} extend the SBOM structure with fields describing component provenance, operational constraints, third-party attestations, regulatory requirements, and organizational policies. These conditions enable users to evaluate the overall trustworthiness of a project based on the reliability and reputation of its components.

\subsubsection{Refinement BOM Framework for Product Lifecycle}
\label{subsubsec:bom_refinement_structural_refinement}
Existing BOM schemas focus mainly on the development stage of a software project and provide limited visibility into later stages like build, deployment, and runtime operation. To address this gap, researchers have proposed lifecycle-aware extensions that add metadata and fields to each operational stage \cite{DRiehleCompute2025, NYousefnezhadBMSD2024, ECornelissenarXiv2025}.
For example, the authors of \cite{DRiehleCompute2025} refined standard SBOM definitions by organizing BOM fields across six lifecycle stages defined by CISA: three supplier-controlled stages (\ie design, source, and build) and three consumer-side stages (``analyzed'' for static code analysis, ``deployed'' for initialization and loading, and ``runtime'' for active execution).
Another work by Yousefnezhad \etal \cite{NYousefnezhadBMSD2024} proposed four BOM classifications aimed at practical DevOps and SecOps needs. Their structure distinguishes four classes: source-code compilation, post-build artifact analysis, runtime monitoring for executed components, and runtime monitoring of host-level network communications.

\subsubsection{Refining BOM Framework for Emerging Digital Ecosystems}
\label{subsubsec:bom_refinement_architecture_adaptation}

Traditional BOM standards were designed for conventional software components, typically assuming single-artifact modules and generic metadata structures. As modern ecosystems, such as multi-tiered digital products and AI-driven systems, grow more complex, these assumptions no longer hold. Therefore, recent studies have introduced domain-specific BOM frameworks or refined existing ones, including adaptions for multi-tier supplier architectures \cite{RBoharaSEW2023} and AI systems \cite{HSidhpurwalaAIMagazine2025, XXuICSAC2025}.

\textbf{Applications with Multi-tiered Suppliers:} 
Modern digital products often rely on multi-tiered suppliers. For example, a software project may combine components from multiple vendors and open-source contributors, which make it hard to generate and maintain a single, centralized BOM. To address this emerging issue, Bohara \etal \cite{RBoharaSEW2023} developed a distributed SBOM (D-SBOM) architecture that allows decentralized generation and storage of SBOM data using blockchain-based systems like IPFS. Because blockchain records are immutable, the resulting SBOMs are traceable and auditable for all stakeholders.

\textbf{AI Systems:}
Existing BOM frameworks struggle to represent the unique components of AI systems, such as model architectures, hyperparameter sets, training datasets, and other AI-specific metadata not captured in traditional software BOMs. To support governance needs, researchers have proposed BOM extensions \cite{HSidhpurwalaAIMagazine2025,XXuICSAC2025} specific to AI systems.
For example, Sidhpurwala \etal \cite{HSidhpurwalaAIMagazine2025} suggested extending AI model cards to include governance-oriented fields such as model intent (\ie purpose and use cases), scope (\ie boundaries and constraints), evaluation data (\eg test sets and performance metrics), governance workflows (\eg approvals and responsible roles), and references to AIBOM safety audits.

\subsection{Proposing New BOM Models for Emerging Challenges} 
\label{subsec:bom_limitation_new_models}

Researchers have also proposed entirely new BOM frameworks to address emerging digital ecosystems that cannot be adequately described by refined versions of existing standards. These new BOM types target domains such as datasets (\S\ref{subsubsec:bom_emerging_data_privacy}), AI systems (\S\ref{subsubsec:bom_emerging_ai_systems}), development and deployment pipelines (\S\ref{subsubsec:bom_emerging_pipeline_automation}), and cyber-physical or IoT systems (\S\ref{subsubsec:bom_emerging_industry_application}).

\subsubsection{Data Provenance and Privacy}
\label{subsubsec:bom_emerging_data_privacy}

High-quality data are critical for modern digital ecosystems, especially AI applications, because the data used for training, validation, and evaluation directly affect their reliability and trustworthiness. At the same time, new regulations such as GDPR, CCPA, EU Data Act increasingly require structured documentation of dataset lineage, usage, and privacy practices. However, traditional BOM standards lack mechanisms to describe the provenance of the dataset or attributes related to privacy, motivating the development of new BOM models (\eg DataBoM, PBOM and PriBOM) \cite{YLiuICB2024, YXiaoSCORED2024, ZTaoPET2025}. 

For example, the authors \cite{YLiuICB2024} proposed DataBOM (Data Bill of Materials) to describe accountability in data supply chains, particularly for AI training data. DataBOM captures detailed metadata for the lifecycle of the dataset, including unique identifiers (\eg UUID, file name, version numbers), licensing information, processing lineage linking derived datasets to their parents, operation history (\eg quality assurance steps, preprocessing actions) and stakeholder identities responsible for the creation or curation of the dataset.
Other works \cite{YXiaoSCORED2024, ZTaoPET2025} extend the BOM concepts to include the privacy-specific properties of data flows. Xiao \etal \cite{YXiaoSCORED2024} proposed the Privacy Bill of Materials (PBOM) to document the privacy attributes of third-party libraries in mobile applications, including collected data types (\eg location and health metrics), data sources, data collection purposes, collection configurations (\eg frequency and conditions), storage mechanisms, and protection indicators. Tao \etal \cite{ZTaoPET2025} introduced PriBOM, another privacy-focused BOM design for generic software applications. PriBOM captures the specifications of the widgets UI (\eg name, type, ID), the backend behaviors (\eg usage of the Android API, data types processed, required permissions), the dependencies of third parties (\eg name, version, release date) and the corresponding privacy policy disclosures.

\subsubsection{AI Systems}
\label{subsubsec:bom_emerging_ai_systems}

Beyond extending existing BOM frameworks for AI systems \S\ref{subsubsec:bom_limitation_study_domain}, researchers also have new BOM frameworks specifically for AI, including AIBOM, DBOM, and NNBoM, which are not yet formally standardized \cite{AMakrisCSR2025, MFazelniaarXiv2024, PRadanlievJDMS2024, JWenzelISoLA2024, QLuCAIN2024, XRenarXiv2025}.

Fazelnia \etal \cite{MFazelniaarXiv2024} proposed AIBOM (Artificial Intelligence Bill of Materials) to document the components, dependencies, and operational context of the AI system. AIBOM is compatible with the AI Vulnerability Database (AIVD) and covers five dimensions: model metadata (\eg name, type, version, dependencies), architectural details (\eg foundation models, software/hardware requirements), data provenance (\eg source, collection method, governance), ethical and environmental considerations, and usage constraints.
Wenzel \etal \cite{JWenzelISoLA2024} introduced the Decision Bill of Materials (DBOM) to enhance decision-level traceability and accountability in AI systems. DBOM records training data, algorithms, runtime environments, and oversight mechanisms.
The authors \cite{XRenarXiv2025} focused on neural networks and proposed NNBoM (Neural Network Bill of Materials) to capture architectural metadata, including pre-trained model invocation patterns, PyTorch module hierarchies (\eg Attention, FeedForward, OnlyMLMHead), and cross-domain co-usage networks that indicate structural reuse across NLP, computer vision, and generative modeling.

\subsubsection{Dynamic Deployment Environments}
\label{subsubsec:bom_emerging_pipeline_automation}

Modern software is increasingly deployed in dynamic environments, such as cloud virtual machines, containerized applications, or infrastructure-as-code setups. These environments can introduce variable software behaviors at runtime, which static BOM frameworks cannot fully capture. Therefore, researchers have developed new BOM frameworks for automated development and deployment pipelines (\eg PBOM and ABOM) \cite{EBandaraIWCMC2025, EBandaraCCNC2025, EBandaraMASS2024, NBoucherarXiv2023}.  

The Pipeline BOM (PBOM) framework \cite{EBandaraIWCMC2025, EBandaraCCNC2025, EBandaraMASS2024} tracks metadata and environment-specific attributes across development and deployment workflows. The main fields include a PBOM identifier (\ie token\_id), generation timestamp, and package information (\ie package\_id and package\_name), and pull request entries that capture environment-specific details such as pull request ID, timestamp, status, contributors, approver, and associated vulnerability ID, name, severity, and fix status.
 Automatic BOM (ABOM) introduced by Boucher \etal \cite{NBoucherarXiv2023} extends this concept by describing component dependencies across different compiler environments (\eg Linux ELF and Windows PE), capturing how deployment conditions can affect component behavior during execution.

\subsubsection{Cyber-physical and IoT Systems}
\label{subsubsec:bom_emerging_industry_application}

Traditional BOM frameworks focus on software components and do not cater for the unique characteristics of critical cyber-physical systems (CPS) and IoT devices, which combine hardware, firmware, networking protocols, and operational behaviors. To address this, recent works propose domain-specific BOM frameworks for critical infrastructure and IoT ecosystems \cite{XYurrebasoarXiv2025, PEmamiNaeiniSSP2020}.

Yurrebaso \etal \cite{XYurrebasoarXiv2025} proposed the Substation Bill of Materials (Subs-BOM), a CycloneDX-based model for digital substations following IEC 61850 standards. Subs-BOM captures device information (\eg vendor, model, hardware revisions), firmware information (\eg versions and  dependencies), exposed IEC 61850 services, and inter-component relationships across substation-device-firmware layers. The authors also extend CycloneDX with a new substation component type that maps IED nameplate attributes from LPHD logical nodes (\eg LPHD.PhyNam.vendor, LPHD.PhyNam.model, LPHD.PhyNam.hwRev) to BOM fields.

For IoT ecosystems, the authors of \cite{PEmamiNaeiniSSP2020} proposed a BOM-like labeling model that documents privacy and security properties of IoT devices. 
The framework uses a two-layer structure: a primary layer showing key user-facing details (\eg data types collected, automatic update policies, default password requirements), a secondary layer containing deeper technical and regulatory information (\eg data retention, encryption methods, third-party sharing, compliance certifications). Both layers are organized into categories such as security controls, data handling practices, and general device attributes to improve transparency for consumers and regulators.

\section{Discussions on Future Research Directions}
\label{sec:bom_future_direction}

After reviewing the current landscape of emerging BOM technologies and their applications, we now outline key research challenges and discuss potential directions to advance BOM generation, representation, and governance across diverse digital technologies.

\subsection{Improving Quality of BOM Data Generation}
\label{subsec:bom_industry_challenges_tool}

Prior research highlighted persistent quality issues in BOM data arising from limitations of current generation tools. These issues include incomplete coverage of dependent components \cite{TBiTSEM2024, LSongarXiv2024, DRiehleCompute2025, SEggersOSTI2023}, inaccurate metadata and component versioning \cite{TBiTSEM2024, TStalnakerICSE2024, SNoceraJSS2025, TNyambeSmartSP2024, ASharmaTSE2025}, inconsistent field representations across tools \cite{SNoceraICSME2024, TBiTSEM2024, DGarciaSVM2025}, and limited support for emerging ecosystems like AI and cyber-physical systems \cite{TNyambeSmartSP2024, TStalnakerICSE2024}.
Those issues largely stem from generation mechanisms that rely mainly on static, text-based information extracted from package managers and source codes, which can miss complex transitive and runtime dependencies \cite{TBiTSEM2024, DGarciaSVM2025} and incomplete modeling of dependency characteristics in emerging digital ecosystems \cite{LSongarXiv2024, ASharmaTSE2025, DRiehleCompute2025, SEggersOSTI2023, TStalnakerICSE2024, TNyambeSmartSP2024, WOtodaSERA2024}.

Future research can improve the quality of BOM data through more dynamic, expressive, and context-aware generation mechanisms. One promising direction is the development of hybrid analysis methodologies that combine static analysis with dynamic behavioral profiling to capture runtime dependency patterns and cross-validate tool outputs. Another important avenue is the design of multi-stage dependency models that distinguish dependency relationships across design-time, build-time, and runtime, enabling more precise tracking of deep transitive chains.
Additionally, domain-specific BOM generation tools for emerging ecosystems (\eg AI systems) are needed to model their unique components (\eg ML models, datasets, hyperparameters) within dependency graphs.

\subsection{Enhancing Trustworthy BOM Data Distribution}
\label{subsec:bom_industry_challenges_sharing}

Previous researchers have identified the absence of secure and standardized distribution channels for BOM data \cite{BXiaICSE2023, GDaliaARES2024}. 
Today, BOMs are often shared using ad-hoc mechanisms, such as email attachments \cite{LJCampSemanticScholar2021} or generic file hosting platforms, which introduce risks of data loss, unauthorized modification, and weak provenance tracking. Even when centralized systems, such as GitHub repositories and package manager systems are used \cite{BXiaICSE2023, GDaliaARES2024}, they remain vulnerable to integrity compromises or inconsistencies between mirrored sources. Therefore, future research should explore standardized dissemination processes and interoperable protocols that incorporate integrity verification and provenance guarantees. Promising directions include employing cryptographic signatures, certificate-based validation, or transparency log mechanisms to ensure resistance against malicious manipulation of generated BOMs. 
Decentralized and tamper-evident distribution architectures, such as blockchain-based registries  \cite{SGuoNOMS2024} may offer trustworthy, community-governed BOM sharing within open-source ecosystems.

\subsection{Guaranteeing Trustability of BOM Data in Production Use}
\label{subsec:bom_industry_challenges_industry}

Previous work has shown that using BOM data in production environments faces practical challenges \cite{TStalnakerICSE2024, SNoceraICSME2024, GDaliaARES2024, BKloegASIACCS2024, LDHolbrookSoftware2023, JLinakerSoftware2023}. A key issue is the limited transparency around BOM data and  the creditability of the tools that generate them \cite{TStalnakerICSE2024, NZahanSP2023, FKCoslettINL2024, GDaliaARES2024}. Currently, there is almost no consistent management process, no certifying authority \cite{LAJaatunCloudCom2023, GDaliaARES2024}, and no systematic quality assurance scheme \cite{AChaoraISTAS2023}, making it difficult for organizations to fully trust and adopt BOM data.  

Future research can help address these barriers by developing community platforms that support the full BOM lifecycle, from trustworthy generation to verifiable distribution and traceable consumption, reducing reliance on today's ad-hoc practices.
Another direction is to design methods to verify the credibility of BOM through third parties without revealing proprietary or sensitive metadata, for example, using differential privacy techniques or certificate-based verification. Researchers may also explore ranking or assessment mechanisms with clear metrics to evaluate the trustworthiness of BOM data.

\subsection{Strengthening Dependency Modeling with Dynamic Behavioral Analysis}
\label{subsec:bom_industry_challenges_emerging_system}
Current BOM frameworks rely almost entirely on static component declarations extracted from manifest files, build configurations, and source code. However, prior work has shown, static analysis alone cannot capture the dynamic, runtime dependencies that emerge during execution and are essential for security monitoring and threat detection \cite{MFazelniaarXiv2024, FKCoslettINL2024, SNoceraICSME2023, NZahanSP2023, NYousefnezhadBMSD2024}.

Future research can address this limitation by incorporating dynamic behavioral analysis into BOM generation and maintenance. By monitoring network activities, function calls and data flows and other runtime behaviors, one can construct dependency models that complement or map to static BOM specifications. Such models capture runtime interactions that static methods inherently miss, thereby enhancing risk assessment and vulnerability analysis. 
Another promising direction is the development of runtime monitoring systems that detect deviations between expected behaviors (as specified by BOM) and the actual behaviors exhibited by deployed components, enabling anomaly detection at runtime.
Finally, systematically integrating behavioral monitoring with static BOM specifications across the full software lifecycle is an important research challenge. Correlating static specifications with observed runtime behaviors can enable automatic discrepancy detection, continuous dependency graph updates, and end-to-end visibility from build-time analysis to pre-deployment validation and operational monitoring.

\section{Related Surveys on Digital Supply Chain Transparency}
\label{sec:bom_related_surveys}

Given the growing importance of transparency and trustworthiness in the complex digital supply chain ecosystem, several survey papers published in the last three years have focused on certain components (\eg software or hardware) and assessment methodologies, ranging from static code analysis to data exchange dependency modeling.

The first cohort of prior work \cite{ACirneCST2024, JPannekampCSUR2024, QLuCSUR2024, ZDengCSUR2025, CMazzoccaCST2025} surveyed the literature on specific digital product (\ie hardware, software, or AI models), without explicitly discussing the notion of a corresponding BOM. A. Cirne \etal \cite{ACirneCST2024} focused on transparency and trust in the hardware of IoT devices through the lens of verifiable proof of identity and chip-embedded resistance to attacks. The authors also touched on issues related to transparency of hardware origins and integrity (\eg traceability of components, verification of authenticity).
The work in \cite{JPannekampCSUR2024} by Pennekamp \etal surveyed existing technologies such as blockchain and provenance systems to provide transparency of information flows between stakeholders (from suppliers to mediators and consumers) and analyzed them in three dimensions: the data being exchanged, the security conditions that govern access and trust, and the utility of information for decision making. 
Lu \etal \cite{QLuCSUR2024} presented a catalog of best practices for the design, development, and governance of AI products, with patterns that support transparency (\eg clear documentation of algorithms and data) and help manage dependencies between AI and non-AI modules. 
Deng \etal \cite{ZDengCSUR2025} proposed a taxonomy for threats to AI agents, highlighting internal risks (\eg reasoning) and interaction risks (\eg engaging with external knowledge sources). They also discussed how mapping agent workflows can help with transparency and dependency challenges.  
Mazzocca \etal \cite{CMazzoccaCST2025} reviewed technologies for decentralized identifiers and verifiable credentials, highlighting their applications in domains such as IoT, healthcare, and transport, and discussing associated security, privacy, and standardization challenges.

The second cohort of survey papers \cite{DGomesCSUR2025, BHammiCSUR2023, BLiTSE2025} focused on assessment methodologies in different aspects of digital ecosystems, from analyzing static code for IoT devices to evaluating operational vulnerabilities in supply chains. Gomes \etal \cite{DGomesCSUR2025} systematically reviewed static code analysis approaches for IoT device security, mapping techniques, and detected vulnerabilities. They highlighted shortcomings such as the heterogeneity of the IoT code and the limited semantic understanding, which lead to false detections. B. Hammi \etal \cite{BHammiCSUR2023} provided a catalog of various types of attacks targeting operational vulnerabilities in digital supply chains (\eg industrial IoT and cyberphysical systems) and mapped the corresponding practical countermeasures. Li \etal \cite{BLiTSE2025} conducted a systematic review of studies on open source software (OSS) license management across three pillars: license identification, risk assessment, and risk mitigation. They highlight challenges such as the absence of detailed metadata models and limited research on compliance.

As a complement to previous surveys, our work focuses on the Bill of Materials (BOM), its data, and associated methods to model and monitor the transparency of a wide variety of digital products, including hardware specifications, software libraries, cloud platforms, cryptographic assets, and AI models. To our knowledge, this is the first paper to systematically survey the development of the BOM standard for digital supply chain transparency, a topic of growing importance supported by industry initiatives and academic research communities.

\section{Conclusion}
\label{sec:bom_survey_conclusion}

This survey provided a comprehensive review of the current industry practices and academic research on Bills of Materials (BOM) for modeling dependencies in digital supply chains, including software (SBOM), hardware (HBOM), cloud services (SaaSBOM) and AI systems (AIBOM). We examined the evolution of BOM standards, from community initiatives to regulatory developments, and summarized the major versions and extensions that address emerging digital ecosystems. By analyzing recent research, we highlighted advances in BOM data generation, security applications, dependency modeling, and structural enhancements to existing frameworks.
Finally, we outlined four promising research directions that can further strengthen the dependency modeling capabilities of BOM and support the reliable operation of modern digital systems.

\bibliographystyle{ACM-Reference-Format}
\bibliography{bomSurvey}

\end{document}